\documentclass[11pt]{article}

\usepackage{amsfonts,amsmath,latexsym,verbatim,amscd,amsthm}

\title{Unique continuation for the vacuum Einstein equations.}

\author{Spyros Alexakis\thanks{Research supported by a Clay Research Fellowhsip.}}
\date{}

\newtheorem{theorem}{Theorem}[section]

\newtheorem{proposition}{Proposition}[section]

\newtheorem{lemma}{Lemma}[section]

\newtheorem{definition}{Definition}[section]

\newcommand{\calC}{{\mathcal C}}

\begin{document}

\maketitle

\abstract{We derive a unique continuation theorem for the vacuum Einstein equations. 
Our method of proof utilizes Carleman estimates
 (most importantly one obtained recently by Ionescu and Klainerman),
 but also relies strongly on certain geometric gauge constructions
 which make it possible to address this problem via such estimates. 
We indicate how our method can be used more broadly to derive unique 
continuation for Einstein's equations from Carleman estimates for the wave operator.}

\section{Introduction.}

\par The main result of this paper is a unique continuation theorem for the vacuum Einstein equations 
across bifurcate horizons. For the reader's convenience we start with 
a rough description of the main theorem: 
 Let $(M,{\bf g}), (\tilde{M},\tilde{\bf g})$
be two vacuum space-times (meaning that $Ric({\bf g})=0, Ric(\tilde{\bf g})=0$), both 
being the global hyperbolic developments 
of space-like hypersurfaces $\Sigma, \tilde{\Sigma}$ respectively. Let $B, \tilde{B}$
be topological open balls in $\Sigma,\tilde{\Sigma}$ and denote by $F, \tilde{F}$ the sets of points
 in $M,\tilde{M}$ respectively that can be joined to  $B,\tilde{B}$ by time-like curves
 (future-directed or past-directed). The regions $F, \tilde{F}$ have boundaries
which we denote by ${\cal H},{\cal \tilde{H}}$. One can visualize ${\cal H}$ 
(resp. ${\cal \tilde{H}}$) as the consisting of the (non-smooth)
 union of two truncated null cones: One truncated cone emanates from $\partial B$ (resp. $\partial \tilde{B}$)
towards the future and the other truncated cone emanates from $\partial B$ (resp. $\partial \tilde{B}$)
towards the past. The region $F$ (resp. $\tilde{F}$) can be thought of as the ``inside'' of ${\cal H}$
(resp. ${\cal \tilde{H}}$); the region $M\setminus F$ (resp. $\tilde{M}\setminus \tilde{F}$) can be thought
of as the ``outside'' of ${\cal H}$ (resp. ${\cal \tilde{H}}$).
 In simple language, we then prove that if the ``insides'' $(F,{\bf g}), (\tilde{F},\tilde{\bf g})$
are isometric then  two open subsets of the ``outsides''
$(M\setminus F,{\bf g}), (\tilde{M}\setminus \tilde{F},\tilde{\bf g})$
must also be isometric.
\newline

\par Unique continuation problems for PDEs have a long history, see \cite{tataru} for
a general discussion. However, such results for
geometric equations (typically equations in the curvature)
have only received attention recently (see \cite{biquard}
where Biquard  derived unique continuation results for Einstein 
metrics of {\it Riemannian} signature).
Such theorems are often proven using Carleman-type estimates and indeed
we will follow this approach in the present paper.
The relevant Carleman estimate that we use comes from a  recent paper of Ionescu and Klainerman
\cite{ik:v1}.

\par In the rest of this introduction we state Theorem \ref{thetheorem} in detail, and 
 make some remarks 
on certain extensions of this result that can be derived by a straightforward modification of 
the proof (see Theorem \ref{easytheorem}). We then briefly
outline some of the arguments in the proof of Theorem \ref{thetheorem}. In section \ref{mainthm} we prove Theorem 
\ref{thetheorem}. For completeness, in section 3 we present a 
derivation of the result of Theorem \ref{thetheorem} under a minimal 
 set of hypotheses. It is worth noting the analogy 
of the calculations in section \ref{power}
with the ones of Rendall in \cite{rendall}.

{\bf The main result:} 
Our theorem deals with vacuum space-times which are maximal developments of incomplete initial data sets:
We will be interested in $\calC^4$ space-times $(M,{\bf g})$ which admit a Caucy hypersurface 
$\Sigma_0\subset M$ where $(\Sigma_0,{\bf g})$ is a ${\cal C}^4$-Riemannian manifold with boundary, 
 $\partial\Sigma_0=S$ is topologically  a 2-sphere and 
 ${\bf g}$ extends in 
a ${\cal C}^4$-fashion to $\partial\Sigma_0$. It follows that in a 
small relatively open neighborhood of $S$, $M$ will have a boundary 
 ${\cal H}$ consisting of the union of 
two null hypersurfaces ${\cal H}^{+}$ and ${\cal H}^{-}$
each of which is ruled by null geodesic rays, so 
that ${\cal H}^{+}$ and ${\cal H}^{-}$ intersect transversely at $S$.
 These two future and past horizons ${\cal H}^{+},{\cal H}^{-}$
are thus each diffeomorphic to $\mathbb{S}^2\times [0,\infty)$ and
the metric ${\bf g}$ restricted to ${\cal H}^{+},{\cal H}^{-}$
is degenerate. Following \cite{ik:v1} we call $S$  the 
{\it bifurcate sphere} and the union ${\cal H}^{+}\bigcup {\cal H}^{-}$
  the {\it bifurcate horizon}.

Our main theorem is then the following:

 \begin{theorem}
\label{thetheorem}
Let $(M,{\bf g})$, $(\tilde{M},{\bf \tilde{g}})$ be two vacuum space-times
($Ric({\bf g})=0$ and $Ric({\bf \tilde{g}})=0$) as described above. Denote by $S,\tilde{S}$
their bifurcate spheres and by ${\cal H}^{+}\bigcup {\cal H}^{-}$, 
${\cal \tilde{H}}^{+}\bigcup {\cal \tilde{H}}^{-}$
their bifurcate  horizons.

 Assume that there exist points $P\in S,\tilde{P}\in\tilde{S}$ and 
relatively open sets $\Omega\subset M,\tilde{\Omega}\subset \tilde{M}$ 
with $P\in \Omega$, $\tilde{P}\in\tilde{\Omega}$ 
containing $S, \tilde{S}$,
 and a diffeomorphism $\Phi: \Omega\rightarrow\tilde{\Omega}$ so that ${\bf g}-\Phi^{*}{\bf \tilde{g}}$
vanishes to third order on $({\cal H}^{+}\bigcup {\cal H}^{-})\bigcap \Omega$.\footnote{By this we mean that
the tensor ${\bf g}-\Phi^{*}{\bf \tilde{g}}$ and all its first and second derivatives
vanish on ${\cal H}^{+}\bigcup {\cal H}^{-}$.}
Then the metrics ${\bf g},{\bf \tilde{g}}$ are isometric in some 
relatively open neighborhoods of $P,\tilde{P}$ in $M,\tilde{M}$. 
 \end{theorem}

{\it Remark 1:} It turns out that the above result can be derived under substantially weaker hypotheses 
on the diffeomorphism $\Phi$: We will show in section \ref{power} that
 it suffices to only assume that the induced conformal structures of the 
horizons ${\cal H}, {\cal \tilde{H}}$
agree near $P,\tilde{P}$, along with certain requirements on the metrics ${\bf g},{\bf \tilde{g}}$ 
restricted to the spheres $S\bigcap \Omega,\tilde{S}\bigcap\tilde{\Omega}$. 
This will essentially follow by a careful analysis of the Einstein 
equations on characteristic hypersurfaces, and is in complete analogy with \cite{rendall}.

{\it Remark 2:} We note that our methods can actually show the following extension of the above:
 Assume that ${\cal H}^{+}, {\cal H}^{-}$ are future/past-complete and also that the metric
 ${\bf g}$ satisfies certain ${\cal C}^1$ bounds near the horizon ${\cal H}^{+}\bigcup {\cal H}^{-}$
(in the interest of brevity we will not make this statement more precise); 
then if there exists a diffeomorphism $\Phi:M\rightarrow \tilde{M}$
for which ${\bf g}-\Phi^{*}{\bf \tilde{g}}$ vanishes to 
third order on the entire horizon ${\cal H}^{+}\bigcup {\cal H}^{-}$,  
the space-times $(M,{\bf g})$, $(\tilde{M},{\bf \tilde{g}})$ will be isometric in open
neighborhoods of the horizons ${\cal H}^{+}\bigcup {\cal H}^{-}$.
We will make a remark further down to point out why this is true.

{\it Remark 3:} In fact, the method we introduce to show Theorem \ref{thetheorem} 
can be applied more widely to show unique continuation across other types of hypersurfaces; 
we indicate how it can be readily adapted to prove unique continuation 
for the vacuum Einstein equations across any smooth time-like hypersurface ${\cal H}$, provided
 H\"ormander's {\it strong pseudo-convexity condition} holds for ${\cal H}$: 
The notion of strong pseudo-convexity is defined for very general classes of operators
 (see the discussion in \cite{tataru}), but for simplicity we will explain 
it only for the wave operator across a smooth time-like surface ${\cal H}$:
 Consider a Lorentzian 
manifold $(M,{\bf g})$ (with an associated wave operator $\Box_g$) 
and a smooth time-like hypersurface ${\cal H}\subset M$ which divides $M$ into regions $M^{+}, M^{-}$. 
We then say that $M^{-}$ is strongly pseudo-convex with
respect to the wave operator $\Box_g$ (or 
equivalently with respect to the metric $g$) near $P\in {\cal H}$
 if there exists an open neighborhood $\Omega$ of $P$ so that  
every {\it null geodesic} in $\Omega$ which is tangent 
to ${\cal H}$ at some point $P'\in \Omega\bigcap {\cal H}$ lies entirely in $M^{+}$, 
 and it only touches ${\cal H}$ at $P'$, with first order of contact.

The next theorem can be proven by a straightforward adaptation of the method of 
proof of Theorem \ref{thetheorem}:

\begin{theorem}
\label{easytheorem} 
Let ${\bf g}, {\bf \tilde{g}}$ be two ${\cal C}^4$ 
Lorentzian metrics defined over a domain
$\Omega\subset \mathbb{R}^4$ satisfying the vacuum Einstein equations: 
$Ric({\bf g})=0, Ric({\bf \tilde{g}})=0$. Let ${\cal H}$ be  a smooth time-like
 hypersurface which divides $\Omega$ into two subdomains 
$\Omega_1,\Omega_2$, and assume that ${\bf g}={\bf \tilde{g}}$ in $\Omega_2$; assume also that 
 $\Omega_1$ satisfies the strong pseudo-convexity condition 
with respect to the metric ${\bf g}$ at $P\in {\cal H}$. 
Then ${\bf g}, {\bf\tilde{g}}$ are isometric in some relatively open neighborhoods of $P$ into $\Omega_1$. 
\end{theorem}

{\it Remark 4:} In the proof of Theorem \ref{thetheorem} we introduce a general method which 
uses Carleman Estimates for the wave operator to derive unique continuation for solutions of 
the vacuum Einstein equations. Thus, Theorem \ref{easytheorem} essentially follows 
by applying this technique 
to the classical Carleman  estimate of H\"ormander 
(see Theorem 4 in \cite{tataru}, or section 28 in \cite{Hormander}
for more details). We will highlight (using separate remarks) 
along the course of the proof of Theorem \ref{thetheorem} the instances where the 
arguments must be slightly altered in order to derive
Theorem \ref{easytheorem}. 
\newline

{\bf Discussion of the Proof of Theorem \ref{thetheorem}:} 
 There are several interesting aspects of applying a Carleman-type estimate
for wave operators to solutions of Einstein's equations in vacuum; in particular,
 the geometric nature of the equations comes starkly into play.

Clearly, in order to reduce the problem to applying a Carleman estimate we must fix a ``canonical''
gauge in which to express the two metrics ${\bf g}, {\bf \tilde{g}}$
and then to use the Einstein equations to derive  a PDE on the {\it difference} of the
 two metrics. The problem is then reduced to
 showing that this difference must vanish in an
 open neighborhood of the bifurcate sphere by applying the Carleman estimate to this PDE.

\par Now, the Ricci curvature is a (non-linear) second order partial
differential operator acting on the metric;
 {\it in wave coordinates} the Ricci curvature has the wave operator
 as its principal symbol. One would therefore
 ideally wish to fix the gauge by  picking wave coordinates
 for ${\bf g}, {\bf \tilde{g}}$ and then subtracting the corresponding
equations $Ric({\bf g})=0, Ric({\bf\tilde{g}})=0$. However this is {\it not possible}: Finding wave
 coordinates in this setting is equivalent to solving
a hyperbolic PDE which is {\it ill-posed} (in the sense that it does not have a solution in general).
Therefore a different choice of gauge must be made, and also a way of circumventing the
fact that the  principle symbol of the Ricci operator will {\it not} be the wave operator must be found.

\par Our remedy to these problems
 is to introduce {\it double Fermi coordinates} and to work with
 a wave equation for the curvature tensors ${\bf R}, {\bf \tilde{R}}$ of the metrics ${\bf g},{\bf \tilde{g}}$:
${\bf \Box_{g}R}={\bf R*R}$,
${\bf \Box_{\tilde{g}}\tilde{R}}={\bf \tilde{R}*\tilde{R}}$.\footnote{These equations follow from
the equation $Ric=0$ via the Bianchi identities.} Double
 Fermi coordinates are constructed by considering a particular null vector field $V$ on ${\cal H}^{+}$
 (obtained through parallel transport along the null generators of ${\cal H}^{+}$)
that points into $M$, and then
constructing the (arc-length parametrized) null geodesics
that emanate from this vector field $V$. This choice of gauge
induces a canonical diffeomorphism $\Psi$ between the
space-times $(M,{\bf g}), (\tilde{M},\tilde{\bf g})$ (locally near $P,\tilde{P}$) which 
reduces the problem to comparing the metrics ${\bf g}, \Psi^{*}{\bf \tilde{g}}$ over $M$.
 We then subtract the two wave equations above
 and derive a wave equation,  (\ref{ss1}), for the {\it difference} ${\bf T_{abcd}}$
of the curvature tensors ${\bf R_{abcd}},{\bf  \tilde{R}_{abcd}}$ of the metrics 
${\bf g}, \Psi^{*}{\bf \tilde{g}}$.
However this equation (\ref{ss1}) also includes terms involving the difference ${\bf d_{ab}}$ of the
two metrics ${\bf g}, \Psi^{*}{\bf \tilde{g}}$, the difference ${\bf G_{ab,c}}$
of their connection coefficients,\footnote{I.e. the difference of their Christoffel symbols
$\Gamma_{ab,c}, \tilde{\Gamma}_{ab,c}$.} and also the derivatives of ${\bf G_{ab,c}}$.

\par The problem then reduces to controlling the weighted $L^2$-norms of these extra
 terms by the weighted $L^2$-norms of the terms ${\bf T}, \partial {\bf T}$ and $\Box {\bf T}$.
Now, {\it in the double Fermi coordinates} we have constructed, the metric is related
to the curvature via an {\it ODE}, (\ref{eqn2}). This allows us to control the weighted $L^2$-norms
of ${\bf d}$ and ${\bf G}$ in (\ref{accent}) by weighted
$L^2$-norms of ${\bf T}$ and $\partial {\bf T}$, which can then
 be absorbed into the Carleman inequality (\ref{1stcarl}).
 However, the equation (\ref{ss1}) also contains certain ``bad terms'' involving {\it derivatives} of
${\bf G_{ab,c}}$; in this setting a straightforward application of
the ODE relation would not allow
us to control the norm of these terms by the norms of the 
terms ${\bf T}, \partial {\bf T}$ and $\Box {\bf T}$.
At this point we make use of the precise algebraic form of the ``bad terms''
(two indices are traced) and another special property of our coordinate system; in particular
{\it in double Fermi  coordinates}, the ``bad terms'' involve no second derivatives of
${\bf d}$ in certain ``bad directions''. This fact, coupled with
standard elliptic estimates on the level sets of the Carleman weight function $f_\epsilon$
 and the
algebraic identities of the curvature tensor, allow us to controll the weighted
$L^2$-norm of the ``bad terms'' by quantities which are allowed in our Carleman estimate.
This enables us to close up the argument and derive that ${\bf T}=0$ (and then ${\bf d}=0$)
 from our Carleman inequality by a standard argument (see \cite{Hormander} or \cite{tataru}).
\newline

\par We now introduce some notational conventions.

{\bf Conventions:} We wish to introduce a dichotomy between
smooth tensor fields defined over $M,\tilde{M}$ and the
components of these smooth tensor fields. We will denote abstract tensor fields with bold letters
e.~g.~ ${\bf A}, {\bf B}$ or if we wish to designate their type or the position of their indices
we will also include the indices: e.~g.~ ${\bf A_\alpha^\beta}$ is a
$(1,1)$-tensor field and ${\bf {B_{\alpha\beta}}^\gamma}$ is a $(1,2)$-tensor field.
On the other hand, once we have constructed a frame field (say $X^0,X^1,X^2,X^3$)
for our manifold below we will denote by $A_a^b$ or ${B_{ab}}^c$ the components of
the above tensor fields with respect to this frame. For example,
$g_{12}$ will stand for the component of the metric tensor (which is
a $(0,2)$-tensor--with lower indices) evaluated for the vectors $X^1,X^2$;
in other words $g_{12}=g(X^1,X^2)$.\footnote{In
fact in most cases below the frame fields we will construct will be the
coordinate vector fields defined by a system of coordinates.} Furthermore, 
throughout the next section, we will have many generic tensor fields appearing as 
coefficients in equations below (e.g. the term $L'^{yu}_{ab}d_{yu}$ 
in (\ref{Beqn2})); unless stated otherwise, 
these will be ${\cal C}^2$-tensor fields over $\Omega$. Also, 
unless we explicitly write out a different summation of indices, repeated 
upper and lower indices (as in 
$L'^{yu}_{ab}$ in (\ref{Beqn2})) will mean that we apply the Einstein summation
 convention and sum over all values ${}^1,{}^2,{}^3,{}^4$, ${}_1,{}_2,{}_3,{}_4$
 that we can give those indices. Finally, we introduce the convention that in section \ref{mainthm}
all estimates involving the parameter $\lambda$ will hold for $\lambda$ large enough, and all constants 
appearing in the estimates will be independent of $\lambda$. 
\newline

 I am grateful to Mihalis Dafermos, Sergiu Klainerman, Alex Ionescu  
and Igor Rodnianski for helpful conversations.

\section{The proof of Theorem \ref{thetheorem}.}
\label{mainthm}

\subsection{Double Fermi coordinates and a PDE-ODE system.}
\label{groundwork}

We will explicitly construct the desired (local, near $P,\tilde{P}$) isometry $\Psi$
between $(M,{\bf g})$ and $(\tilde{M},{\bf \tilde{g}})$. 
In order to do this, we firstly construct a useful set of 
coordinates outside the bifurcate horizon ${\cal H}^{+}\bigcup {\cal H}^{-}$.

Consider the sphere $S$ in $(M,{\bf g})$ and let $\Omega$ be 
a small relatively open neighborhood 
of $P\in S$; pick a pair of null vector fields $U,V$
on $S$ with the following two properties: Firstly, $U$ is future-directed
and tangent to ${\cal H}^{+}$ and $V$ is past-directed and tangent to ${\cal H}^{-}$. Secondly,
$g(U,V)= 1$ on all of $S\bigcap \Omega$. Now, consider the affine-parametrized null geodesics emanating from $U$
(these will correspond to the null generators of ${\cal H}^{+}$);
for each $A\in S$ we denote by $l_A$ the null geodesic that thus emanates from $P$.
Notice that given any coordinate system defined on 
$S\bigcap \Omega$,  ${\cal Y}:S\bigcap \Omega\rightarrow \mathbb{R}^2$,
we obtain a coordinate system ${\cal Y}':{\cal H}^{+}\bigcap \Omega\rightarrow \mathbb{R}^2\times [0,1)$.

\par Next, we parallel-transport the vectors $V$ along the null geodesics $l_A$:
Thus, for each point $Q\in {\cal H}^{+}\bigcap \Omega$  we obtain
a past-directed outward pointing null vector $V_Q$. Finally,
consider the (affine-parametrized) null geodesics emanating from the vectors $V_Q$.
 We have thus obtained a coordinate system of the form
 ${\cal Y}'':\Omega'\rightarrow \mathbb{R}^2\times [0,1)\times[0,\delta_0)$, where
$\Omega'$ is some relatively open neighborhood of $P$ in the 
space-time $M$.\footnote{By slight abuse of notation we will denote $\Omega'$ by $\Omega$ again.}

\begin{definition}
\label{doubferm}

For any space-time $(M,{\bf g})$ as in the hypothesis of Theorem \ref{thetheorem}
we call a system of coordinates
as above ``double Fermi coordinates''.
\end{definition}

\par We now consider the vector fields $\tilde{U},\tilde{V}$ in $\tilde{M}$ where
$\tilde{U}=\Phi_{*}U, \tilde{V}=\Phi_{*}V$ and also the coordinate system
 $\tilde{\cal Y}={\cal Y}\circ \Phi^{-1}:\tilde{S}\bigcap \tilde{\Omega}\rightarrow \mathbb{R}^2$.
 We perform the same construction as above (for $\delta_0$ small enough) 
for the space-time $\tilde{M}$, obtaining a new coordinate system
$\tilde{\cal Y}:\tilde{\Omega}\rightarrow \mathbb{R}^2\times [0,1)\times [0,\delta_0)$,
 where $\tilde{\Omega}$ is a relatively open neighborhood of $\tilde{P}$ 
 in the space-time $\tilde{M}$.
Consider the map 
 $\Psi:\Omega\rightarrow \tilde{\Omega}$ defined by the formula:

$$\Psi=\tilde{\cal Y}^{-1}\circ {\cal Y}.$$

\par Let us pull back the metric ${\bf \tilde{g}}$ to $M$ via this map: We
define ${\bf g}'=\Psi^{*}{\bf \tilde{g}}$. We will show that ${\bf g'}={\bf g}$ in
an open neighborhood of $P\in S$. That will prove our claim. 
\newline

Two remarks are in order here: Firstly, in view of the freedom of picking the vector 
fields $U,V$ over $S$ (in particular since we are only imposing the requirement $g(U,V)=1$, 
we could just as well replace these vector fields by $\tau U,\frac{1}{\tau}V$) our argument 
below can be used to show that if there is a map $\Phi:M\longrightarrow \tilde{M}$ for which 
${\bf g}-\Phi^{*}{\bf \tilde{g}}$ vanishes up to second order on
 all of ${\cal H}^{+}\bigcup {\cal H}^{-}$, then 
$(M,{\bf g})$, $(\tilde{M},{\bf \tilde{g}})$ are isometric in an open neighborhood of the 
whole horizon ${\cal H}^{+}\bigcup {\cal H}^{-}$--the 
${\cal C}^1$-bounds on ${\bf g}$ mentioned in Remark 2 serve to ensure that the 
constant $\epsilon$ below  can be picked independently of $\tau$. 
\newline

Now, we wish to study the components of the metrics ${\bf g},{\bf g'}$ with respect to the
coordinate system over $M$ that we have constructed.

Let $x^1,x^2$ be  coordinate functions on $S$, defined near $P\in S$ such that $x^1(P)=x^2(P)=0$; 
let $x^3$ be the coordinate
on ${\cal H}^{+}$ defined by the null geodesics emanating from the vectors $U_P$ through the
equation $\nabla_{\frac{\partial}{\partial x^3}}\frac{\partial}{\partial x^3}=0$.
Thus we have coordinates $x^1,x^2,x^3$ defined on ${\cal H}^{+}$, near $P$. Now consider
the coordinate $x^0$ in $\Omega$ defined by the null geodesics emanating in the direction of
$V_Q$ through the
equation $\nabla_{\frac{\partial}{\partial x^0}}\frac{\partial}{\partial x^0}=0$.
Thus we obtain coordinates $x^0,x^1,x^2,x^3$ in the open set $\Omega$.

Consider the metric $g_{ab}$ (where the lower indices
can take values ${}_0,{}_1,{}_2,{}_3$ that correspond to the above coordinate system). Given the equation
$\nabla_{\frac{\partial}{\partial x^0}}\frac{\partial}{\partial x^0}=0$
we derive the equations: $\partial_0 g_{0a}=0$ for ${}_a={}_0,{}_1,{}_2,{}_3$.
Thus for every point in $\Omega$ we get: $g_{00}=0,g_{01}=0,g_{02}=0, g_{03}=1$.
Analogously we derive that $g'_{00}=0,g'_{01}=0,g'_{02}=0, g'_{03}=1$.

\par Now, denote by $\Gamma_{ab,c}=\frac{1}{2}[\partial_ag_{bc}+\partial_bg_{ac}-\partial_cg_{ab}]$,
 $\Gamma'_{ab,c}=\frac{1}{2}[\partial_ag'_{bc}+\partial_bg'_{ac}-\partial_cg'_{ab}]$
the Chiristoffel symbols of the metrics $g,g'$ in the coordinates of $M$
that we have constructed. Consider also the curvature tensors $R_{abcd}, R'_{abcd}$
(with 4 lower indices) of the metrics $g,g'$ and also the Levi-Civita
connections $\nabla,\nabla'$ of the metrics $g,g'$. We define the tensors ${\bf d,T,G,D}$
through the equations:\footnote{Note that a tensor is specified by specifying its components
relative to a frame.}

$$d_{ab}=g_{ab}-g'_{ab},
T_{abcd}=R_{abcd}-R'_{abcd},
G_{ab,c}=\Gamma_{ab,c}-\Gamma'_{ab,c}, D_{ab}=\partial_sG_{ta,b}g^{st}.$$

We will prove that in some open neighborhood $\Omega'$ of $S$,
${\bf T}_{abcd}=0$ and ${\bf d}_{ab}=0$. That will prove our theorem.
\newline

Our next goal is to derive a system of equations (both PDEs and ODEs) in the tensors above.

Consider the components $R_{0ab0}, R'_{0ab0}$ of the curvature tensors for $g,g'$.
By the definition of curvature tensor we derive that {\it in the double Fermi coordinates}:

\begin{equation}
 \label{eqn2}
R_{0ab0}=\frac{1}{2}\partial^{(2)}_{00}g_{ab}+\frac{1}{4}\partial_0g_{as}\partial_0g_{tb}g^{st},
\end{equation}

\begin{equation}
 \label{eqn2'}
R'_{0ab0}=\frac{1}{2}\partial^{(2)}_{00}g'_{ab}+\frac{1}{4}\partial_0g'_{as}\partial_0g'_{tb}g'^{st}.
\end{equation}

Subtracting the above two equations we derive:

\begin{equation}
\label{Beqn2}
T_{0ab0}=\frac{1}{2}\partial^{(2)}_{00}d_{ab}+L_{ab}^{yu}\partial_0d_{yu}+L'^{yu}_{ab}d_{yu}.
\end{equation}

\par Now, consider the two equations:

\begin{equation}
\label{curvwaves}
\bf{ \Box_gR=R*R}, {\bf \Box_{g'}R'=R'*R'}.
\end{equation}

We are going to subtract these two equations. We introduce some notation first;
We let $\widetilde{\Box_g}$ be the ``rough wave operator'':
$g^{ab}\partial_a\partial_b$ which acts on scalar-valued functions.
We also note that
$g^{ab}-g'^{ab}=-g^{as}d_{st}g'^{tb}.$
Then, subtracting the two equations in (\ref{curvwaves}) 
we derive an equation which hold for any values ${}_a,{}_b,{}_c,{}_d={}_0,{}_1,{}_2,{}_3$:

$$\widetilde{\Box_g} T_{abcd}= F_{abcd}^{yuvx}T_{yuvx}+F'^{yuvxb}_{abcd}\partial_bT_{yuvx}
+F''^{yu}_{abcd}d_{yu}+F'''^{yuv}_{abcd}\partial_v d_{yu}+F''''^{yu}_{abcd}D_{yu}.$$
(here the tensor fields $F'',F'''$ are $\calC^1,\calC^0$ respectively).

To derive the next set of equations, we will break the tensor ${\bf D_{ab}}$ into the
symmetric part ${\bf D_{(ab)}}$ (${\bf D_{(ab)}}=\frac{1}{2}[{\bf D_{ab}+D_{ba}}]$) and  the antisymmetric part
${\bf D_{[ab]}}$ (${\bf D_{[ab]}}=\frac{1}{2}[{\bf D_{ab}-D_{ba}}]$). By the definition of the Christoffel symbols
we see that $D_{(ab)}$ equals:
$D_{(ab)}=-\frac{1}{2}\widetilde{\Box_g}d_{ab}$.

On the other hand we calculate: $D_{[ab]}=\frac{1}{2}[\partial^{(2)}_{sa}d_{tb}-\partial^{(2)}_{sb}d_{sa}]g^{st}$.
Now, in order to derive an equation on $D_{[ab]}$ we consider the equations:
$\nabla^sR_{0sab}=0$, $\nabla'^sR'_{0sab}=0$ and we subtract them. We derive an equation:

\begin{equation}
\label{eqn4}
g^{st}[\partial_{s0a}d_{tb}-\partial_{s0b}d_{ta}]= C^{yu}_{ab}d_{yu}+
C'^{yuv}_{ab}\partial_yd_{uv}+C''^{yuvx}_{ab}T_{yuvx}+C'''^{yu}_{ab}D_{yu}
+C''''^{yuv}\partial_{0v}d_{yu}.
\end{equation}

In fact, we observe that because of the form of the metric $g_{ab}$
(with lower indices) we must have $g^{3a}=g^{a3}=0$ for $a=1,2,3$
and $g^{03}=g^{30}=1$, therefore the above gives us an equation:

\begin{equation}
\label{eqn4'}
\partial_0D_{[ab]}= C^{yu}_{ab}d_{yu}+
C'^{yuv}_{ab}\partial_yd_{uv}+C_{ab}^{yuvx}T_{yuvx}+C'''^{yu}_{ab}D_{yu}
+C''''^{yuv}_{ab}\partial_{0v}d_{yu}+\sum_{i,j=1}^2c^{ij}\partial_iG_{j[a,b]}.
\end{equation}

Thus our system of equations is as follows:

\begin{eqnarray}
\label{ss1}
&\widetilde{\Box_g} T_{abcd}= F_{abcd}^{yuvx}T_{yuvx}+F^{yuvxb}_{abcd}\partial_b
T_{yuvx}
\\&\notag+F^{yu}_{abcd}d_{yu}+F^{yuv}_{abcd}\partial_yd_{uv}+F^{yu}_{abcd}D_{yu},
\label{ss2}
\\&T_{0ab0}=\frac{1}{2}\partial^{(2)}_{00}d_{ab}+L_{ab}^{yu}\partial_0d_{yu}+L'^{yu}_{ab}d_{yu},
\label{ss3}
\\& G_{ab,c}=\frac{1}{2}[\partial_ad_{bc}+\partial_bd_{ac}-\partial_cd_{ab}], D_{ab}=g^{st}\partial_sG_{ta,b},
\label{ss4}
\\& D_{(ab)}=-\frac{1}{2}\widetilde{\Box_g}d_{ab},
\label{ss5}
\\&\partial_0D_{[ab]}= C^{yu}_{ab}d_{yu}+
C^{yuv}_{ab}\partial_yd_{uv}
\\&\notag +C_{ab}^{yuvx}T_{yuvx}+C'^{yu}_{ab}D_{yu}+
C'^{yuv}_{ab}\partial^{(2)}_{0v}d_{yu}+\sum_{i,j=1}^2c^{ij}\partial_iG_{j[a,b]}.
\end{eqnarray}

{\it Remark 5:} The analogue of the double Fermi coordinates in the setting of Theorem \ref{easytheorem}
is as follows: 
In this case we can pick  a hypersurface $S\subset \Omega_2$ which touches $\cal H$
 to first order at $P$, and which is still strongly pseudo-convex. We then 
 (locally near $P\in S$) pick coordinates $x^1,x^2,x^3$ on $S$, 
 such that $x^3$ is a  time-like direction and $x^1,x^2$ are space-like. Finally, we 
  let $\vec{\nu}$ be the (space-like) unit normal vector field to $S$, which points towards $\Omega_1$ and 
consider the arc-length parametrized space-like geodesics that emanate from $\vec{\nu}$. 
For each of the metrics ${\bf g}, {\bf \tilde{g}}$, this defines 
a fourth coordinate function $x^0$ 
in a relatively open neighborhood $\tilde{\Omega}$ of $P$, on the side of $S$ that intersects $\Omega_1$. We have  
thus obtained a ``canonical coordinate system'' for both metrics ${\bf g},{\bf \tilde{g}}$.
 We then construct the map $\Psi:\tilde{\Omega}\rightarrow \tilde{\Omega}$ by identifying the coordinates 
for the two metrics ${\bf g}, {\bf \tilde{g}}$. By hypothesis 
we know that $\Psi$ fixes $S$ and $\Psi^{*}{\bf \tilde{g}}-{\bf g}$
vanishes to third order on $S$ (since ${\bf g}={\bf \tilde{g}}$ in $\Omega_2$).
They key feature is that we 
 have that $g_{0i}=g'_{0i}=0$ for $i=1,2,3$ and $g_{00}=g'_{00}=1$ in $\tilde{\Omega}$.  
This allows us to derive the same system of equations as above. We note that as in  \cite{tataru}, 
the function $x^0$ will be strongly pseudo-convex
in a small enough neighborhood of $P$.

\subsection{The Ionescu-Klainerman Carleman Estimate and our Main Proposition.}

\par To state our main Proposition we must recall some results from \cite{ik:v1}.
The reader is referred to section 6 in that paper. Firstly
recall the optical functions $u_{+},u_{-}$ defined near $S$.
We condider the coordinate system $\{\frac{u_{+}+u_{-}}{\sqrt{2}},\frac{u_+-u_{-}}{\sqrt{2}}, x^1,x^2 \}$ 
defined over $\Omega$ and the function $N^P$ defined over $\Omega$ via 
$N(P):=[u_{+}^2+u_{-}^2+(x^1)^2+(x^2)^2]$;\footnote{The function 
$N^P$ is defined to be a distance function with respect to a Euclidean coordinate system 
around $P$, in the exterior region. For our purposes, we  choose the coordinate system 
$\{\frac{u_{+}+u_{-}}{\sqrt{2}},\frac{u_+-u_{-}}{\sqrt{2}},x^1,x^2 \}$ such that 
 $N(P):=u_{+}^2+u_{-}^2+(x^1)^2+(x^2)^2$.} recall also that 
that $B_{\epsilon^{10}}(P)$ stands for the set of points in $\Omega$ for which $(N^P)^2\le \epsilon^{20}$. 
Then the Carleman weight function of Ionescu-Klainerman is:
$f_\epsilon=ln\{\epsilon^{-1}(u_{+}+\epsilon)(u_{-}+\epsilon)+\epsilon^{10}N^{P}\}$ 
(defined for some fixed small $\epsilon>0$) from Lemma 6.2 in \cite{ik:v1}.

 For any scalar-valued function $\Phi$ defined
 on $B_{\epsilon^{10}}$ we recall the weighted $L^2$-norm introduced in \cite{ik:v1}:

$$||\Phi||_{L^2_\lambda}=\sqrt{\int_{B_{\epsilon^{10}}}
\Phi^2 e^{-2\lambda\cdot f_\epsilon} dV_g}.$$

\par We introduce a cut-off function $\chi:\mathbb{R}\rightarrow [0,1]$,
 defined to be smooth and supported in $[1/2,\infty]$, 
and equal to 1 in $[3/4,\infty]$. We then define $\eta_\epsilon:M\rightarrow [0,1]$
 to be $\eta_\epsilon(x):=1-\chi(N^P/\epsilon^{10}(x))$. (In the setting 
 of Theorem \ref{easytheorem} we can pick any cut-off function
  $\eta_\epsilon(x^0)=1-\chi(\frac{x^0}{\epsilon})$, 
for $\epsilon>0$ small enough). 

We  denote by $V_{\epsilon}$ a generic
function (independent of $\lambda$) which is supported in the set
where $0<\eta_{\epsilon}<1$. We then recall the first Carleman estimate of
Ionescu and Klainerman (see Lemma 6.2 in \cite{ik:v1}):
Setting $\phi=T_{abcd}\cdot \eta_{\epsilon}$ we derive that
there exist a constants $\epsilon,C,\lambda_0$ so that
for every $\lambda>\lambda_0$:

\begin{equation}
\label{1stcarl}
\begin{split}
&\lambda\cdot \sum_{a,b,c,d=0}^3||T_{abcd}\cdot \eta_{\epsilon}||_{L^2_\lambda}+
\sum_{a,b,c,d,e=0}^3||\partial_e T_{abcd}\cdot \eta_{\epsilon}||_{L^2_\lambda}
\le
\\&\frac{C}{\sqrt{\lambda}} \{\sum_{a,b,c,d=0}^3
||\widetilde{\Box_g}T_{abcd}\cdot \eta_{\epsilon}||_{L^2_\lambda}+
||V_{\epsilon}||_{L^2_\lambda}\}.
\end{split}
\end{equation}
{\it Remark 7:} In \cite{ik:v1} this estimate is derived for functions $\phi$
 in $C_0^\infty (B_{\epsilon^{10}})$; in fact,  following the proof of this estimate in \cite{ik:v1}
 we observe that that it holds for ${\cal C}^2$-functions $\phi$ which 
 vanish on $\partial B_{\epsilon^{10}}\bigcap ({\cal H}^{+}\bigcup {\cal H}^{-})$
 and vanish along with their first derivatives on 
 $\partial B_{\epsilon^{10}}\setminus (\partial B_{\epsilon^{0}}\bigcap ({\cal H}^{+}\bigcup {\cal H}^{-})$. 
 Thus we are allowed to set $\phi=T_{abcd}\cdot \eta_{\epsilon}$ and derive (\ref{1stcarl}).

{\it Remark 8:} In the setting of Theorem \ref{easytheorem} the analogous Carleman estimate for functions which are compactly supported in a 
small enough neighborhood of $P$ in $\Omega_1$ is classical, see \cite{tataru}; in that setting
 the weight function can be chosen to be $f_\epsilon:=x^0$, and the estimate holds for compactly supported 
 functions in a neighborhood of $P$ where the level sets of $x^0$ are strongly pseudo-convex, 
and for $\epsilon>0$ small enough so that all the intersections 
$\{x^0=\epsilon'\}\bigcap \Omega_1$, $\epsilon'<\epsilon$, are compact. 
\newline

Now, using the equation (\ref{ss1}) we derive that there exists a 
constant $C'$ (independent of $\lambda$) so that:

\begin{equation}
 \label{accent}
\begin{split}
&||\widetilde{\Box_g}T_{abcd}\cdot \eta_{\epsilon}||_{L^2_\lambda}
\le C' \{\sum_{a,b,c,d=0}^3||T_{abcd}||_{L^2_\lambda}
+\sum_{a,b,c,d,e=0}^3||\partial_eT_{abcd}||_{L^2_\lambda}
\\&+\sum_{a,b=0}^3||d_{ab}||_{L^2_\lambda}+
\sum_{a,b,c=0}^3||\partial_c d_{ab}||_{L^2_\lambda}
\sum_{a,b=0}^3+||D_{ab}||_{L^2_\lambda}+
||V_{\epsilon}||_{L^2_\lambda}\}.
\end{split}
\end{equation}

Our main Proposition is the following:

\begin{proposition}
\label{theprop}
We claim that there exists a (universal) constant $C$ and a number $\lambda_0>0$
so that for every $\lambda\ge \lambda_0$:

\begin{equation}
\label{theeqn1}
\sum_{a,b=0}^3||d_{ab}||_{L^2_\lambda}\le \frac{C}{\sqrt{\lambda}}\{
\sum_{a,b,c,d=0}^3||T_{abcd}\cdot \eta_{\epsilon}||_{L^2_\lambda}
+||V_{\epsilon}||_{L^2_\lambda}\},
\end{equation}

\begin{equation}
\label{theeqn2}
\sum_{a,b,c=0}^3||\partial_cd_{ab}||_{L^2_\lambda}\le \frac{C}{\sqrt{\lambda}}\{
\sum_{a,b,c,d=0}^3||T_{abcd}\cdot \eta_{\epsilon}||_{L^2_\lambda}
+\sum_{a,b,c,d,e=0}^3||\partial_eT_{abcd}\cdot \eta_{\epsilon}||_{L^2_\lambda}
+||V_{\epsilon}||_{L^2_\lambda}\},
\end{equation}

\begin{equation}
\label{theeqn3}
\begin{split}
&\sum_{a,b=0}^3||D_{ab}||_{L^2_\lambda}\le \frac{C}{\sqrt{\lambda}}\{
\sum_{a,b,c,d=0}^3||T_{abcd}\cdot \eta_{\epsilon}||_{L^2_\lambda}
+\sum_{a,b,c,d,e=0}^3||\partial_eT_{abcd}\cdot \eta_{\epsilon}||_{L^2_\lambda}
\\&+\sum_{a,b,c,d=0}^3||\widetilde{\Box_g}T_{abcd}\cdot \eta_\epsilon||_{L^2_\lambda}
+||V_{\epsilon}||_{L^2_\lambda}\}.
\end{split}
\end{equation}
\end{proposition}

Let us check how the above will imply Theorem \ref{thetheorem}: Using the above
three estimates and  (\ref{accent})
we derive that there exists a $\lambda_0>0$ and a constant $C'$ independent of $\lambda$ 
so that for every $\lambda>\lambda_0$:

\begin{equation}
 \label{accent'}
\begin{split}
&||\widetilde{\Box_g}T_{abcd}\cdot \eta_{\epsilon}||_{L^2_\lambda}\le C'
 \{\sum_{a,b,c,d=0}^3||T_{abcd}||_{L^2_\lambda}
+\sum_{a,b,c,d,e=0}^3||\partial_eT_{abcd}||_{L^2_\lambda}
+||V_{\epsilon}||_{L^2_\lambda}\}.
\end{split}
\end{equation}
Thus, replacing the above into (\ref{1stcarl}) we derive that for $\lambda$ large enough:

\begin{equation}
\label{1stcarlconc}
\begin{split}
&\lambda\cdot \sum_{a,b,c,d=0}^3||T_{abcd}\cdot
 \eta_{\epsilon}||_{L^2_\lambda}+
\sum_{a,b,c,d,e=0}^3||\partial_e T_{abcd}\cdot
\eta_{\epsilon}||_{L^2_\lambda}\le
\frac{C}{\sqrt{\lambda}}
||V_{\epsilon}||_{L^2_\lambda}.
\end{split}
\end{equation}
Now, the argument from page 35 in \cite{ik:v1}  implies that 
 $T_{abcd}=0$ for all $\{{}_a,{}_b,{}_c,{}_d\}\in \{{}_0,{}_1,{}_2,{}_3,\}$ and for every
$P\in B_{\epsilon^{40}}$.\footnote{The point here is that the
 maximum value of the weight function $e^{-\lambda f_\epsilon}$ 
 in the support of $V_\epsilon$ is bounded 
above by minimum  value of $e^{-\lambda f_\epsilon}$ in $B_{\epsilon^{40}}$.} 
Then, using (\ref{ss3}) we derive that $d_{ab}=0$
for all $\{{}_a,{}_b\}\in \{{}_0,{}_1,{}_2,{}_3,\}$ in $ B_{\epsilon^{40}}$.
This shows that ${\bf g}={\bf g'}$ in $B_{\epsilon^{40}}$. Similarly,
 in the setting of Theorem \ref{easytheorem}
we derive that $d_{ab}=0$ in the region $x^0\le \frac{\epsilon}{2}$. 
\newline

\subsection{Proof of Proposition \ref{theprop}:}

We now prove a main Lemma which will be very useful towards proving Proposition
\ref{theprop}.
Firstly let us make a note regarding the relation between the coordinate $x^0$
and the function $u_{+}$ introduced in \cite{ik:v1}. Recall that ${\cal H}^{+}=\{x^0=0\}=\{u_{+}=0\}$. 
Moreover there is a function $\rho$ defined over
$B_{\epsilon^{10}}$ so that for
every point $P\in B_{\epsilon^{10}}$:
$\frac{\partial}{\partial u_{+}}=\rho(P)\frac{\partial}{\partial x^0}$.
There clearly exist numbers $0<\mu\le M$ so that for every $P\in B_{\epsilon^{10}}$,
$0<\mu\le \rho(P)\le M$. We now state our main claim:

\begin{lemma}
\label{derivlem}
Let $\phi$ be a function defined in $B_{\epsilon^{10}}$
which vanishes on $B_{\epsilon^{10}}\bigcap {\cal H}^{+}$.\footnote{Note that
 we are not requiring $\phi$ to be compactly supported
in $B_{\epsilon^{10}}$.} Then we claim there exists a $C>0$ so that for $\lambda$ large enough:

\begin{equation}
\label{thebound}
||e^{-\lambda\cdot f_\epsilon} \phi||_{L^2(B_{\epsilon^{10}})}\le 
\frac{C}{\sqrt{\lambda}}||e^{-\lambda\cdot f_\epsilon} \partial_0\phi||_{L^2(B_{\epsilon^{10}})}.
\end{equation}
\end{lemma}

{\it Proof:} Firstly a note about the volume form: 
The volume form $dV_g$ is defined by a function $\omega(u_{+},x^1,x^2,u_{-})$ 
defined over $B_{\epsilon^{10}}$ so that:

$$dV_g= \omega(u_{+},x^1,x^2,u_{-}) du_{+}\wedge dx^1\wedge dx^2 \wedge du_{-}.$$
Note that there exists constants $\mu',M'$ so that 
 $0<\mu'\le\omega(P)\le M'$ for every $P\in B_{\epsilon^{10}}$. 
By definition:

$$||e^{-\lambda\cdot f_\epsilon} \phi||^2_{L^2(B_{\epsilon^{10}})}=
\int_{B_{\epsilon^{10}}} e^{-2\lambda\cdot f_\epsilon}\phi^2 \omega
du_{+}\wedge dx^1\wedge dx^2\wedge du_{-}.$$
We observe that $e^{-2\lambda f_\epsilon}=[\epsilon^{-1}(u_{+}+\epsilon)(u_{-}+\epsilon)
+\epsilon^{10}(u_{+}^2+u_{-}^2+(x^1)^2+(x^2)^2)]^{-2\lambda}$.
Given fixed values for $u_{-},x^1,x^2$, we set 
 $w_\epsilon(t):=[\epsilon^{-1}(t+\epsilon)(u_{-}+\epsilon)
+\epsilon^{10}(t^2+u_{-}^2+(x^1)^2+(x^2)^2)]$.\footnote{I.e. we are allowing
the parameter $u_{+}$ to vary, and label it $s$.} Observe that for $(t, u_-,x^1,x^2)\in  B_{\epsilon^{10}}$ 
we have a bound:
$$1\le \partial_{t}w_\epsilon(t)\le 1+2\epsilon.$$
Now, given fixed vaues for  $x^1,x^2,u_{-}$ such that $(u_-)^2+(x^1)^2+(x^2)^2\le \epsilon^{20}$, we let 
$Max(u_-,x^1,x^2):=\sqrt{\epsilon^{20}-(u_-)^2-(x^1)^2-(x^2)^2}$; we claim:

$$\int_0^{Max(u_-,x^1,x^2)} (w_\epsilon(u_{+}))^{-2\lambda}\phi^2 \omega
du_{+}\le \frac{C}{\lambda}\int_0^{Max(u_-,x^1,x^2)}
(w_\epsilon(u_{+}))^{-2\lambda}(\partial_0 \phi)^2 \omega
du_{+},$$
with the constant $C$ independent of $x^1,x^2,u_-$.
Clearly this will imply our claim. 

We now prove the above. Some notational conventions: We write $Max$ instead of $Max(u_-,x^1,x^2)$ for short. 
Moreover, when we
write $\partial_{u_{+}}$ we will be referring to
differentiation with respect to the vector field $\frac{\partial}{\partial u_{+}}$,
while $\partial_0$ will stand for differentiation
with respect to the vector field $\frac{\partial}{\partial x^0}$. We  derive:

\begin{equation}
\label{theest}
\begin{split}
&\int_0^{Max} (w_\epsilon(u_{+}))^{-2\lambda}\phi^2\omega
du_{+}\le M\int_0^{Max} (w_\epsilon(u_{+}))^{-2\lambda}\phi^2
du_{+}=\int_0^{Max} (w_\epsilon(u_{+}))^{-2\lambda}(\int_0^{u_{+}}\partial_{u_{+}}\phi dt)^2
du_{+}
\\&\le M\int_0^{Max} [(w_\epsilon(u_{+}))^{-2\lambda}(\int_0^{u_{+}}
(w_\epsilon(t))^{-2\lambda}(\partial_{u_{+}}\phi)^2 dt)(\int_0^{u_{+}}
(w_\epsilon(t))^{2\lambda}dt)]du_{+}\le
\\& M'[\int_0^{Max} (w_\epsilon(u_{+}))^{-2\lambda}(\partial_{u_{+}} \phi)^2 du_{+}]
\cdot \int_0^{Max} [(w_\epsilon(u_{+}))^{-2\lambda}
\int_0^{u_{+}}(w_\epsilon(t))^{2\lambda}(\partial_tw_\epsilon(t)) dt]du_{+}\le
\\&M'[\int_0^{Max} (w_\epsilon(u_{+}))^{-2\lambda}(\partial_{u_{+}} \phi)^2 
du_{+}]\cdot\int_0^{Max} \frac{w_\epsilon(u_+)}{2\lambda+1}du_{+}\le
\\&\frac{M''}{\lambda}\int_0^{Max} (u_{+}+\epsilon)^{-2\lambda}(\partial_{u_{+}} \phi)^2 du_{+}\le
\frac{C}{\lambda}\int_0^{Max}
(u_{+}+\epsilon)^{-2\lambda}(\partial_0 \phi)^2 \omega du_{+},
\end{split}
\end{equation}
QED. $\Box$

{\it Remark 9:} In the setting of Theorem \ref{easytheorem} we can derive 
the exact same Lemma (with a gain of a factor $\frac{C}{\sqrt{\lambda}}$ in the RHS),
 with the classical weight
 function $e^{-\lambda \psi(x)}$,\footnote{Here $\psi(x)=x^0$, in the notation of remark 5.}
in a small enough neighborhood of the point $P$.

\par We now use the above result to derive some estimates:

\begin{lemma}
\label{easylem}
We claim that there exist constants $C,\lambda_0$ so that for every $\lambda>\lambda_0$:

\begin{equation}
\label{easyeqn1}
\sum_{a,b=0}^3||d_{ab}\cdot\eta_{\epsilon}||_{L^2_\lambda}
\le \frac{C}{\sqrt{\lambda}}\{\sum_{a,b,c,d=0}^3||T_{abcd}\cdot
\eta_{\epsilon}||_{L^2_\lambda}+||V_{\epsilon} ||_{L^2_\lambda}\},
\end{equation}

\begin{equation}
\label{easyeqn2}
\sum_{a,b,c=0}^3||\partial_c d_{ab}\cdot\eta_{\epsilon}||_{L^2_\lambda}
\le \frac{C}{\sqrt{\lambda}}\{\sum_{a,b,c,d=0}^3||T_{abcd}\cdot\eta_{\epsilon}||_{L^2_\lambda}
+\sum_{a,b,c,d,e=0}^3||\partial_eT_{abcd}\cdot\eta_{\epsilon}
||_{L^2_\lambda}+||V_{\epsilon} ||_{L^2_\lambda}\},
\end{equation}

\begin{equation}
\label{easyeqn3}
\sum_{a,b,c=0}^3||\partial^{(2)}_{0c} d_{ab}\cdot\eta_{\epsilon}||_{L^2_\lambda}
\le \frac{C}{\sqrt{\lambda}}\{\sum_{a,b,c,d=0}^3||T_{abcd}\cdot\eta_{\epsilon}||_{L^2_\lambda}
+\sum_{a,b,c,d,e=0}^3||\partial_eT_{abcd}\cdot\eta_{\epsilon}
||_{L^2_\lambda}+||V_{\epsilon} ||_{L^2_\lambda}\},
\end{equation}

\begin{equation}
\label{easyeqn4}
\sum_{a,b,c=0}^3||\partial^{(3)}_{00c} d_{ab}\cdot\eta_{\epsilon}||_{L^2_\lambda}
\le \{\sum_{a,b,c,d=0}^3||T_{abcd}\cdot\eta_{\epsilon}||_{L^2_\lambda}+
\sum_{a,b,c,d,e=0}^3||\partial_eT_{abcd}\cdot\eta_{\epsilon}
||_{L^2_\lambda}+||V_{\epsilon} ||_{L^2_\lambda}\}.
\end{equation}
\end{lemma}

{\it Proof :} We will prove (\ref{easyeqn1}). The other equations hold by the same argument. 
To prove (\ref{easyeqn1}) we 
repeatedly use the Lemma \ref{derivlem}. By applying it once we derive:

\begin{equation}
\label{tsoxatz}
\begin{split}
&\sum_{a,b=0}^3||d_{ab}\cdot\eta_{\epsilon} ||_{L^2_\lambda}
\le \frac{C}{\sqrt{\lambda}}\sum_{a,b=0}^3||\partial_0(d_{ab}\cdot\eta_{\epsilon}) ||_{L^2_\lambda}\le 
\\&\frac{C}{\sqrt{\lambda}}\sum_{a,b=0}^3\{||(\partial_0d_{ab})\cdot\eta_{\epsilon} ||_{L^2_\lambda}+
 \frac{C}{\sqrt{\lambda}}\sum_{a,b=0}^3||d_{ab}\cdot\partial_0(\eta_{\epsilon}) ||_{L^2_\lambda}\}.
\end{split}
\end{equation}

Analogously we derive:

\begin{equation}
\label{tsoxatz2}
\begin{split}
&\sum_{a,b=0}^3||(\partial_0d_{ab})\cdot\eta_{\epsilon} ||_{L^2_\lambda}
\le \frac{C}{\sqrt{\lambda}}\sum_{a,b=0}^3||\partial_0[(\partial_0d_{ab})\cdot\eta_{\epsilon}] ||_{L^2_\lambda}\le 
\\&\frac{C}{\sqrt{\lambda}}\sum_{a,b=0}^3\{||(\partial^{(2)}_{00}d_{ab})
\cdot\eta_{\epsilon} ||_{L^2_\lambda}+ 
\sum_{a,b=0}^3||\partial_0d_{ab}\cdot\partial_0(\eta_{\epsilon}) ||_{L^2_\lambda}\}\le
\\&\frac{C}{\sqrt{\lambda}}\sum_{a,b=0}^3\{||2(T_{0ab0}
-L^{yu}_{ab}\partial_0d_{yu}-L'^{yu}_{ab}d_{yu})\cdot\eta_{\epsilon} ||_{L^2_\lambda}+ \sum_{a,b=0}^3||V_{\epsilon} ||_{L^2_\lambda}\}\le
\\&\frac{C}{\sqrt{\lambda}}\sum_{a,b=0}^3\{||T_{0ab0}\cdot\eta_{\epsilon} ||_{L^2_\lambda}+
\sum_{a,b=0}^3||\partial_0d_{ab}\cdot\eta_{\epsilon} ||_{L^2_\lambda}+
\sum_{a,b=0}^3||d_{ab}\cdot\eta_{\epsilon} ||_{L^2_\lambda}+
||V_{\epsilon} ||_{L^2_\lambda}\}.
\end{split}
\end{equation}

Now, we can control the term $\sum_{a,b=0}^3||d_{ab}\cdot\eta_{\epsilon}
||_{L^2_\lambda}$ in the RHS using (\ref{tsoxatz});
furthermore, for $\lambda$ large enough the term $\frac{C}{\sqrt{\lambda}}
\sum_{a,b=0}^3||\partial_0d_{ab}\cdot
\eta_{\epsilon} ||_{L^2_\lambda}$ in
the RHS can be absorbed into the LHS and thus we derive the equation:

\begin{equation}
\label{tsoxatz3'}
\begin{split}
&\sum_{a,b=0}^3||(\partial_0d_{ab})\cdot\eta_{\epsilon} ||_{L^2_\lambda}
\le\frac{C}{\sqrt{\lambda}}\{\sum_{a,b=0}^3||T_{0ab0}\cdot\eta_{\epsilon} ||_{L^2_\lambda}
+\sum_{a,b=0}^3||V_{\epsilon} ||_{L^2_\lambda}\}.
\end{split}
\end{equation}

Now, combining (\ref{tsoxatz3'}) with (\ref{tsoxatz})
we also derive (\ref{easyeqn1}). Equations (\ref{easyeqn2}), (\ref{easyeqn3}), (\ref{easyeqn4})
follow by a straightforward adaptation of this argument.  $\Box$
\newline

\par We now claim a more involved Lemma.

\begin{lemma}
\label{hardlamma}
We claim that:

\begin{equation}
\label{suid}
\begin{split}
&\sum_{a,b,c=0}^3||D_{ab}\cdot\eta_{\epsilon}
||_{L^2_\lambda}\le \frac{C}{\sqrt{\lambda}}
\{\sum_{a,b,c,d=0}^3||T_{abcd}\cdot\eta_{\epsilon}
||_{L^2_\lambda}+\sum_{a,b,c,d,e=0}^3||\partial_eT_{abcd}
\cdot\eta_{\epsilon}
||_{L^2_\lambda}+
\\&\sum_{a,b,c,d=0}^3||\widetilde{\Box_g}T_{abcd}\cdot\eta_{\epsilon}
||_{L^2_\lambda}
+||V_{\epsilon}||_{L^2_\lambda}\}.
\end{split}
\end{equation}
\end{lemma}

Note that 
in view of Lemma \ref{easylem}, this Lemma will imply the main Proposition \ref{theprop}, 
and hence also Theorem \ref{thetheorem}.
\newline

{\it Proof of Lemma \ref{hardlamma}:} We will prove the above in two pieces. Firstly, recall that
$\bf{D}_{(ab)}$ stands for the symmetric part of the 2-tensor $\bf{D}_{ab}$ (so $\bf{D}_{(ab)}=\frac{1}{2}[\bf{D}_{ab}+\bf{D}_{ba}]$) and
$\bf{D}_{[ab]}$ stands for the antisymmetric part
(so $\bf{D}_{[ab]}=\frac{1}{2}[\bf{D}_{ab}-\bf{D}_{ba}]$). We will prove (\ref{suid})
separately for the symmetric part $\bf{D}_{(ab)}$ and the
antisymmetric part $\bf{D}_{[ab]}$ of $\bf{D}_{ab}$. Specifically we will prove:

\begin{equation}
\label{suid1}
\begin{split}
&\sum_{a,b=0}^3||D_{(ab)}\cdot\eta_{\epsilon}
||_{L^2_\lambda}\le \frac{C}{\sqrt{\lambda}}
\{\sum_{a,b,c,d=0}^3||T_{abcd}\cdot\eta_{\epsilon}
||_{L^2_\lambda}+\sum_{a,b,c,d,e=0}^3||\partial_eT_{abcd}
\cdot\eta_{\epsilon}
||_{L^2_\lambda}
\\&+\sum_{a,b,c,d=0}^3||\widetilde{\Box_g}T_{abcd}\cdot\eta_{\epsilon}
||_{L^2_\lambda}
+||V_{\epsilon}||_{L^2_\lambda}\},
\end{split}
\end{equation}

\begin{equation}
\label{suid2}
\begin{split}
&\sum_{a,b=0}^3||D_{[ab]}\cdot\eta_{\epsilon}
||_{L^2_\lambda}\le \frac{C}{\sqrt{\lambda}}
\{\sum_{a,b,c,d=0}^3||T_{abcd}\cdot\eta_{\epsilon}
||_{L^2_\lambda}+\sum_{a,b,c,d,e=0}^3||\partial_eT_{abcd}
\cdot\eta_{\epsilon}
||_{L^2_\lambda}
\\&+\sum_{a,b,c,d=0}^3||\widetilde{\Box_g}T_{abcd}\cdot\eta_{\epsilon}
||_{L^2_\lambda}
+||V_{\epsilon}||_{L^2_\lambda}\}.
\end{split}
\end{equation}

{\bf Proof of (\ref{suid1}):} Recall that
 $D_{(ab)}=-\frac{1}{2}\widetilde{\Box_g}d_{ab}(=-\frac{1}{2}g^{yu}\partial^{(2)}_{yu}d_{ab}$).
Let us also make an important observation: Since we have $g_{0b}=g_{b0}=0$
for $b=0,1,2$ and $g_{03}=g_{30}=1$, we derive that $g^{3b}=g^{b3}=0$ for $b=0,1,2$
and $g^{03}=g^{30}=1$. We now claim two useful estimates which we will prove later.
First useful estimate:

\begin{equation}
\label{stim1}
\begin{split}
&\sum_{a,b=0}^3||(\partial_0g^{yu})\partial^{(2)}_{yu}d_{ab}\cdot\eta_{\epsilon}
||_{L^2_\lambda}\le
C\{\sum_{a,b=0}^3||\widetilde{\Box_g}d_{ab}\cdot\eta_{\epsilon}
||_{L^2_\lambda}+\sum_{a,b,c,d=0}^3||T_{abcd}\cdot\eta_{\epsilon}
||_{L^2_\lambda}
\\&+\sum_{a,b,c,d,e=0}^3||\partial_eT_{abcd}
\cdot\eta_{\epsilon}
||_{L^2_\lambda}+
||V_{\epsilon}||_{L^2_\lambda}\}.
\end{split}
\end{equation}

Second estimate:

\begin{equation}
\label{stim2}
\begin{split}
&\sum_{a,b=0}^3||(\partial_0g^{yu})\partial^{(3)}_{yu0}d_{ab}\cdot\eta_{\epsilon}
||_{L^2_\lambda}\le C
\{\sum_{a,b=0}^3||\widetilde{\Box_g}\partial_0d_{ab}\cdot\eta_{\epsilon}
||_{L^2_\lambda}+\sum_{a,b,c,d=0}^3||T_{abcd}\cdot\eta_{\epsilon}
||_{L^2_\lambda}
\\&+\sum_{a,b,c,d,e=0}^3||\partial_eT_{abcd}
\cdot\eta_{\epsilon}
||_{L^2_\lambda}+
||V_{\epsilon}||_{L^2_\lambda}\}.
\end{split}
\end{equation}

Let us check how the two equations (\ref{stim1}), (\ref{stim2}) will
imply (\ref{suid1}). We start by applying Lemma \ref{derivlem}:

\begin{equation}
\label{fw1}
\begin{split}
&\sum_{a,b=0}^3||\widetilde{\Box_g}d_{ab}\cdot\eta_{\epsilon}
||_{L^2_\lambda}\le \frac{C}{\sqrt{\lambda}}
[\sum_{a,b=0}^3||\partial_0[\widetilde{\Box_g}d_{ab}\cdot\eta_{\epsilon} ]
||_{L^2_\lambda}]\le
\\&\frac{C}{\sqrt{\lambda}}
\{\sum_{a,b=0}^3||g^{yu}\partial^{(3)}_{yu0}d_{ab}\cdot\eta_{\epsilon}
||_{L^2_\lambda}+ \sum_{a,b=0}^3||(\partial_0g^{yu})\partial^{(2)}_{yu}d_{ab}\cdot\eta_{\epsilon}
||_{L^2_\lambda}+
||V_{\epsilon}||_{L^2_\lambda}\}.
\end{split}
\end{equation}

Therefore, using (\ref{stim1}) and taking $\lambda$ large enough we derive:

\begin{equation}
\label{fw2}
\begin{split}
&\sum_{a,b=0}^3||\widetilde{\Box_g}d_{ab}\cdot\eta_{\epsilon}
||_{L^2_\lambda}\le
\frac{C}{\sqrt{\lambda}}
\{\sum_{a,b=0}^3||g^{yu}\partial^{(3)}_{yu0}d_{ab}\cdot\eta_{\epsilon}
||_{L^2_\lambda}
\\&+\sum_{a,b,c,d=0}^3||T_{abcd}\cdot\eta_{\epsilon}
||_{L^2_\lambda}
+\sum_{a,b,c,d,e=0}^3||\partial_eT_{abcd}
\cdot\eta_{\epsilon}
||_{L^2_\lambda}+
||V_{\epsilon}||_{L^2_\lambda}\}.
\end{split}
\end{equation}

 Again applying Lemma \ref{derivlem} we derive:

\begin{equation}
\label{fw3}
\begin{split}
&\sum_{a,b=0}^3||g^{yu}\partial^{(3)}_{yu0}d_{ab}\cdot\eta_{\epsilon}
||_{L^2_\lambda}\le \frac{C}{\sqrt{\lambda}}
\sum_{a,b=0}^3||\partial_0[g^{yu}\partial^{(3)}_{yu0}d_{ab}\cdot\eta_{\epsilon}
||_{L^2_\lambda}]\le
\\&\frac{C}{\sqrt{\lambda}}\{\sum_{a,b=0}^3||g^{yu}\partial^{(4)}_{yu00}d_{ab}\cdot\eta_{\epsilon}
||_{L^2_\lambda}+
\sum_{a,b=0}^3||(\partial_0g^{yu})\partial^{(3)}_{yu0}d_{ab}\cdot\eta_{\epsilon}
||_{L^2_\lambda}+
||V_{\epsilon}||_{L^2_\lambda}\}\le
\\&\frac{C}{\sqrt{\lambda}}\{\sum_{a,b=0}^3||2g^{yu}\partial^{(2)}_{yu}(T_{0ab0}-
L^{yu}_{ab}\partial_0d_{yu}-L'^{yu}_{ab}d_{yu})\cdot\eta_{\epsilon}||_{L^2_\lambda}+
\\&\sum_{a,b=0}^3||(\partial_0g^{yu})\partial^{(3)}_{yu0}d_{ab}\cdot\eta_{\epsilon}
||_{L^2_\lambda}+
||V_{\epsilon}||_{L^2_\lambda}\}\le
\frac{C}{\sqrt{\lambda}}\{\sum_{a,b=0}^3||2g^{yu}\partial^{(2)}_{yu}
T_{0ab0}\cdot\eta_{\epsilon}
||_{L^2_\lambda}+
\\&\sum_{a,b=0}^3||g^{yu}\partial^{(3)}_{yu0}d_{ab}\cdot\eta_{\epsilon}
||_{L^2_\lambda}+
\sum_{a,b=0}^3||g^{yu}\partial^{(2)}_{yu}d_{ab}\cdot\eta_{\epsilon}
||_{L^2_\lambda}+
\\&\sum_{a,b,c=0}^3||\partial^{(2)}_{0c}d_{ab}\cdot\eta_{\epsilon}
||_{L^2_\lambda}+
\sum_{a,b,c=0}^3||\partial_{c}d_{ab}\cdot\eta_{\epsilon}
||_{L^2_\lambda}+
\sum_{a,b=0}^3||d_{ab}\cdot\eta_{\epsilon}
||_{L^2_\lambda}+
\\&\sum_{a,b=0}^3||(\partial_0g^{yu})\partial^{(3)}_{yu0}d_{ab}\cdot\eta_{\epsilon}
||_{L^2_\lambda}+
||V_{\epsilon}||_{L^2_\lambda}\}.
\end{split}
\end{equation}

Now, the terms $\sum_{a,b,c=0}^3||\partial^{(2)}_{0c}d_{ab}\cdot\eta_{\epsilon}
||_{L^2_\lambda}$,
$\sum_{a,b,c=0}^3||\partial_{c}d_{ab}\cdot\eta_{\epsilon}
||_{L^2_\lambda}$,
$\sum_{a,b=0}^3||d_{ab}\cdot\eta_{\epsilon}
||_{L^2_\lambda}$
can be controlled by virtue of Lemma \ref{easylem}.
Then, combining (\ref{fw1}) and (\ref{fw3}) to also control the term $\sum_{a,b=0}^3||g^{yu}\partial^{(2)}_{yu}d_{ab}\cdot\eta_{\epsilon}
||_{L^2_\lambda}$ ($=\sum_{a,b=0}^3||\widetilde{\Box_g}d_{ab}\cdot\eta_{\epsilon}
||_{L^2_\lambda}$) and absorb it into the LHS we derive:

\begin{equation}
\label{fw4}
\begin{split}
&\sum_{a,b=0}^3||g^{yu}\partial^{(3)}_{yu0}d_{ab}\cdot\eta_{\epsilon}
||_{L^2_\lambda}\le
\frac{C}{\sqrt{\lambda}}\{\sum_{a,b=0}^3||g^{yu}\partial^{(3)}_{yu0}d_{ab}\cdot\eta_{\epsilon}
||_{L^2_\lambda}+
\sum_{a,b=0}^3||\widetilde{\Box_g}
T_{0ab0}\cdot\eta_{\epsilon}
||_{L^2_\lambda}
\\&+\sum_{a,b,c,d=0}^3||T_{abcd}\cdot\eta_{\epsilon}
||_{L^2_\lambda}
+\sum_{a,b,c,d,e=0}^3||\partial_eT_{abcd}
\cdot\eta_{\epsilon}
||_{L^2_\lambda}+
||V_{\epsilon}||_{L^2_\lambda}\}.
\end{split}
\end{equation}
 Therefore for $\lambda$ large enough we can absorb the term
$||g^{yu}\partial^{(3)}_{yu0}d_{ab}\cdot\eta_{\epsilon}||_{L^2_\lambda}$
from the RHS into the LHS; then, substituting in this estimate into the (\ref{fw2}) we derive 
 (\ref{suid1}),
subject to proving (\ref{stim1}), (\ref{stim2}). We now prove these two equations:
\newline

{\bf Proof of (\ref{stim1}):} The key in the proof of this estimate is
the fact that $\partial_0 g^{3b}=0$ for $b=0,1,2,3$,\footnote{This remains true 
in the setting of Theorem \ref{easytheorem}.} 
and that  we have already controlled the weighted $L^2$-norms of the
functions $\partial^{(2)}_{0c}d_{ab}$. With these observations the
desired estimate (\ref{stim1}) will readily be reduced to elliptic estimates on the level sets of the
Carleman weight function $f_\epsilon$:
Since $g^{3b}=0$ for $b=1,2,3$, we derive that:

$$|(\partial_0g^{cd})\partial^{(2)}_{cd}d_{ab}|^2\le \sum_{c,d=1}^2|(\partial_0g^{cd})
\partial^{(2)}_{cd}d_{ab}|^2+2\sum_{r=0}^3|(\partial_0g^{r0})\partial^{(2)}_{0r}d_{ab}|^2.$$

Let us make a few notes that will be useful further down.
Firstly recall that 
 the Fermi coordinate system we have introduced defines the coordinate vector fields:
Let $\{X^0,X^1,X^2,X^3\}$ be the vector fields $\{\frac{\partial}{\partial x^0}, 
\frac{\partial}{\partial x^1},\frac{\partial}{\partial x^2},\frac{\partial}{\partial x^3}\}$.
Then  when we
give values ${}_0,{}_1,{}_2,{}_3$ to the lower
indices ${}_\alpha,{}_\beta$ in $\partial_{\alpha\beta}$ those values
correspond to these vector fields $X^0,X^1,X^2,X^3$ above. In that notation, recall the ``rough wave operator'':

\begin{equation}
\label{rough1}
\widetilde{\Box_g}=\sum_{c,d=1}^2g^{cd}\partial^{(2)}_{cd}+
2\sum_{c=1}^3 g^{0r}\partial^{(2)}_{0r}+g^{00}\partial^{(2)}_{00}.
\end{equation}

We will consider the operator $\sum_{c,d=1}^2g^{cd}\partial^{(2)}_{cd}$ separately and
denote it by $\tilde{\Delta}_{g,1}$; we call it the ``first rough Laplacian''.
We will now introduce a second basis for $T(B_{\epsilon^{10}})$ which will be useful:
Recall the Carleman weight function $f_\epsilon$. By virtue of the form of $f_\epsilon$ 
and the fact that $\{x^3=C\}=\{ u_{-}=C\}$, 
we derive that there exist smooth functions $\rho^1,\rho^2$ defined 
over $B_{\epsilon^{10}}$ such that 
the vector fields $\tilde{X}^1=X^1+\rho^1X^0$ and $\tilde{X}^2=X^2+\rho^2X^0$ are
tangent to the level sets of $f_\epsilon$.\footnote{In the setting of Theorem \ref{easytheorem}
 there is no need to modify the vector fieilds $X^1,X^2$--these are
already tangent to the level sets of the Carleman weight function.} Thus, we obtain a new frame 
at each point $P\in B_{\epsilon^{10}}$,
$\{\tilde{X}^0=X^0,\tilde{X}^1,\tilde{X}^2,\tilde{X}^3=X^3\}$. When we
refer to components of tensors with respect to this 
new frame we will use indices with tildes e.g. 
${}_{\tilde{0}},{}^{\tilde{2}}$ etc. We observe that
$g_{\tilde{0}\tilde{a}}= g_{\tilde{a}\tilde{0}}=0$ for ${}_a={}_0,{}_1,{}_2$ and also
$g_{\tilde{0}\tilde{3}}= g_{\tilde{3}\tilde{0}}=1$ thus again we derive that
$g^{\tilde{3}\tilde{a}}=g^{\tilde{a}\tilde{3}}=0$ for
${}^{\tilde{a}}={}^{\tilde{0}},{}^{\tilde{1}},{}^{\tilde{2}}$ and
$g^{\tilde{3}\tilde{0}}=g^{\tilde{0}\tilde{3}}=1$. In fact we observe that
$g^{\tilde{c}\tilde{d}}=g^{cd}$ for ${}^c,{}^d={}^1,{}^2$.
We now define the ``second
rough wave operator'' with respect to this frame:

\begin{equation}
\label{rough2}\widetilde{\Box_{g,2}}=\sum_{\tilde{c},\tilde{d}=1}^2g^{\tilde{c}\tilde{d}}
\partial^{(2)}_{\tilde{c}\tilde{d}}
+\sum_{\tilde{r}=\tilde{1}}^3 g^{\tilde{0}\tilde{r}}\partial^{(2)}_{\tilde{0}\tilde{r}}+
g^{\tilde{0}\tilde{0}}\partial^{(2)}_{\tilde{0}\tilde{0}}.
\end{equation}

Observe that:

\begin{equation}
\label{rough3}\widetilde{\Box_g}=\widetilde{\Box_{g,2}}+\sum_{i=0}^3L^i\partial_i.
\end{equation}
(So the 2nd term in the RHS is a generic linear combination of first order operators with $\calC^3$ coefficients).
We again separately consider the operator $\sum_{\tilde{c},\tilde{d}=1}^2g^{\tilde{c}\tilde{d}}\partial^{(2)}_{\tilde{c}\tilde{d}}$
which we will call the {\it second rough Laplacian}
and denote it by $\tilde{\Delta}_{g,2}$. Observe that:

\begin{equation}
\label{rough4}
\tilde{\Delta}_g=\tilde{\Delta}_{g,2}+\sum_{i=0}^2 R^i\partial_i+\sum_{i=0}^2 R'^i\partial^{(2)}_{0i}.
\end{equation}

It then follows straightforwardly that there exists a constant $C>0$ such that:

\begin{equation}
\label{conclnrough}
[\sum_{c,d=1}^2|(\partial_0g^{cd})
\partial^{(2)}_{cd}d_{ab} |]^2 \le C [\sum_{\tilde{c},\tilde{d},\tilde{e},\tilde{r}=1}^2 g^{\tilde{c}\tilde{d}}g^{\tilde{e}\tilde{r}}\partial^{(2)}_{\tilde{c}\tilde{e}}d_{ab}
\partial^{(2)}_{\tilde{d}\tilde{r}}d_{ab}
+\sum_{c=0}^2[(\partial_cd_{ab})^2+
(\partial^{(2)}_{0c}d_{ab})^2]].
\end{equation}

Using the above we can now prove (\ref{stim1}). Directly applying the above we obtain:

\begin{equation}
\label{prfstim1}
\begin{split}
&\sum_{a,b=0}^3||(\partial_0g^{yu})\partial^{(2)}_{yu}d_{ab}\cdot\eta_{\epsilon}
||^2_{L^2_\lambda}\le C'\cdot \sum_{c,d=1}^2 \int_{B_{\epsilon^{10}}}[g^{\tilde{c}\tilde{d}}g^{\tilde{e}\tilde{r}}
\partial^{(2)}_{\tilde{c}\tilde{e}}d_{ab}
\partial^{(2)}_{\tilde{d}\tilde{r}}d_{ab}]\cdot e^{-2\lambda f_\epsilon}\eta^2_{\epsilon}dV_g
\\&+\sum_{c=0}^3||\partial_cd_{ab}\cdot\eta_{\epsilon}||^2_{L^2_\lambda}+
\sum_{c=0}^3||\partial^{(2)}_{0c}d_{ab}\cdot\eta_{\epsilon}||^2_{L^2_\lambda}.
\end{split}
\end{equation}
(Here $dV_g$ is the 
volume form for the metric $g$ on $B_{\epsilon^{10}}$).

Now, the trick is to estimate the term $\sum_{c,d=1}^2 \int_{B_{\epsilon^{10}}}
[g^{\tilde{c}\tilde{d}}g^{\tilde{e}\tilde{r}}\partial^{(2)}_{\tilde{c}\tilde{e}}d_{ab}
\partial^{(2)}_{\tilde{d}\tilde{r}}d_{ab}]\cdot
e^{-2\lambda f_\epsilon}\eta_{\epsilon}dV_g$
in the first line of the RHS.
We  integrate by parts twice with respect to $\partial_{\tilde{d}}$
and then $\partial_{\tilde{e}}$; since the vector fields 
$\tilde{X}^1,\tilde{X}^2$ (with respect to which we are integrating by 
parts) are {\it tangent} to the level sets of $f_\epsilon$, 
we will {\it not} bring out derivatives of the factor $e^{-2\lambda f_\epsilon}$.  We derive:

\begin{equation}
\label{karaman1}
\begin{split}
&\sum_{\tilde{c},\tilde{d},\tilde{e},\tilde{r}=1}^{\tilde{2}} \int_{B_{\epsilon^{10}}}
[g^{\tilde{c}\tilde{d}}g^{\tilde{e}\tilde{r}}\partial^{(2)}_{\tilde{c}\tilde{e}}d_{ab}
\partial^{(2)}_{\tilde{d}\tilde{r}}d_{ab}]\cdot
e^{-2\lambda f_\epsilon}\cdot\eta^2_{\epsilon} dV_g=
\\&\int_{B_{\epsilon^{10}}}
[\tilde{\Delta}_{g,2}d_{ab}]^2\cdot e^{-2\lambda f_\epsilon}\eta^2_{\epsilon}dV_g
+\int_{B_{\epsilon^{10}}}
L^{yu}\partial_yd_{ab}\partial_ud_{ab}\cdot e^{-2\lambda f_\epsilon}\eta^2_{\epsilon}
dV_g+ 
\\&\int_{B_{\epsilon^{10}}} L^y\partial_yd_{ab}
\tilde{\Delta}_{g,2}d_{ab}\cdot e^{-2\lambda f_\epsilon}\eta^2_{\epsilon}dV_g
+\int_{B_{\epsilon^{10}}}
V_{\epsilon}e^{-2\lambda f_\epsilon}
\eta^2_{\epsilon}dV_g.
\end{split}
\end{equation}

Thus applying Cauchy-Schwartz to the above we deduce:

\begin{equation}
\label{prfstim2}
\begin{split}
&\sum_{\tilde{c},\tilde{d},\tilde{e},\tilde{r}=1}^{\tilde{2}} \int_{B_{\epsilon^{10}}}
[g^{\tilde{c}\tilde{d}}g^{\tilde{e}\tilde{r}}\partial^{(2)}_{\tilde{c}\tilde{e}}d_{ab}
\partial^{(2)}_{\tilde{d}\tilde{r}}d_{ab}]\cdot
e^{-2\lambda f_\epsilon}\eta^2_{\epsilon}dV_g\le C\{\int_{B_{\epsilon^{10}}}
[\tilde{\Delta}_{g,2}d_{ab}]^2\cdot e^{-2\lambda f_\epsilon}\eta^2_{\epsilon}dV_g
\\&+\sum_{c=0}^2\int_{B_{\epsilon^{10}}}(\partial_cd_{ab})^2
 e^{-2\lambda f_\epsilon}\eta^2_{\epsilon}dV_g
+\int_{S_{u_{+},u_{-}}}V_{\epsilon}e^{-2\lambda f_\epsilon}
\eta^2_{\epsilon}dV_g.
\end{split}
\end{equation}

 Replacing the above into (\ref{prfstim1}) we derive:

 \begin{equation}
\label{prfstim3}
\begin{split}
&\sum_{a,b=0}^3||(\partial_0g^{yu})\partial^{(2)}_{yu}d_{ab}\cdot\eta_{\epsilon}
||_{L^2_\lambda}\le C\{\sum_{a,b=0}^3|| \tilde{\Delta}_{g,2}d_{ab}\cdot \eta_{\epsilon}||_{L^2_\lambda}+
\\& \sum_{c=0}^3||\partial^{(2)}_{c0}d_{ab} ||_{L^2_\lambda} +\sum_{c=0}^3||\partial_cd_{ab} ||_{L^2_\lambda}
+||V_{\epsilon}||_{L^2_\lambda}\}.
\end{split}
\end{equation}

Now, using the formulas (\ref{rough1})--(\ref{rough4}) we 
derive that
$| \tilde{\Delta}_{g,2}d_{ab}|=|\widetilde{\Box_g}d_{ab} +\sum_{r=0}^3C^r\partial^{(2)}_{0r}d_{ab}
+\sum_{r=0}^3C'^r\partial_{r}d_{ab}|$
for some $\calC^2$ functions $C^r,C'^r$. Thus substituting the above
into  (\ref{prfstim3}) we derive that:

 \begin{equation}
\label{prfstim3}
\begin{split}
&\sum_{a,b=0}^3||(\partial_0g^{yu})\partial^{(2)}_{yu}d_{ab}\cdot\eta_{\epsilon}
||_{L^2_\lambda}\le C\{\sum_{a,b=0}^3|| \widetilde{\Box_g}d_{ab}\cdot \eta_{\epsilon}||_{L^2_\lambda}+
\\& \sum_{c=0}^3||\partial^{(2)}_{c0}d_{ab} ||_{L^2_\lambda} +\sum_{c=0}^3||\partial_cd_{ab}||_{L^2_\lambda}+
||V_{\epsilon}||_{L^2_\lambda}\}.
\end{split}
\end{equation}

Thus, combining the above with Lemma \ref{easylem} we derive our claim.
\newline

{\bf Proof of (\ref{stim2}):}
The proof of this claim is very much in the spirit of the previous one.
 We again use the formulas (\ref{rough1})--(\ref{rough4})
to derive the analogue of (\ref{conclnrough}):

\begin{equation}
\label{conclnrough'}
[\sum_{c,d=1}^2|(\partial_0g^{cd})
\partial^{(3)}_{cd0}d_{ab} |]^2 \le C \{
\sum_{\tilde{c},\tilde{d},\tilde{e},\tilde{r}=\tilde{1}}^{\tilde{2}} g^{\tilde{c}\tilde{d}}g^{\tilde{e}\tilde{r}}\partial^{(3)}_{\tilde{c}\tilde{e}\tilde{0}}d_{ab}
\partial^{(3)}_{\tilde{d}\tilde{r}\tilde{0}}d_{ab}
+\sum_{c=0}^3[(\partial_cd_{ab})^2+(\partial^{(2)}_{c0}d_{ab})^2
(\partial^{(3)}_{c00}d_{ab})^2]\}.
\end{equation}

\par Thus, using the above we derive an analogue of  (\ref{prfstim1}):

\begin{equation}
\label{prfstim1'}
\begin{split}
&\sum_{a,b=0}^3||(\partial_0g^{yu})\partial^{(3)}_{yu0}d_{ab}\cdot\eta_{\epsilon}
||^2_{L^2_\lambda}\le C'\cdot \{\sum_{\tilde{c},\tilde{d},\tilde{e},\tilde{r}=
\tilde{1}}^{\tilde{2}} \int_{B_{\epsilon^{10}}}[g^{\tilde{c}\tilde{d}}g^{\tilde{e}\tilde{r}}
\partial^{(3)}_{\tilde{c}\tilde{e}\tilde{0}}d_{ab}
\partial^{(3)}_{\tilde{d}\tilde{r}\tilde{0}}d_{ab}]\cdot e^{-2\lambda f_\epsilon}
\eta^2_{\epsilon}dV_g
\\&+\sum_{c=0}^3[||\partial^{(2)}_{c0}d_{ab}
\cdot\eta_{\epsilon}||^2_{L^2_\lambda}+
||\partial^{(3)}_{c00}d_{ab}\cdot\eta_{\epsilon}||^2_{L^2_\lambda}+
||\partial^{(2)}_{c0}d_{ab}\cdot\eta_{\epsilon}||^2_{L^2_\lambda}]\}=
\\&C'\cdot \sum_{c,d=1}^2 \int_{B_{\epsilon^{10}}}
[g^{\tilde{c}\tilde{d}}g^{\tilde{e}\tilde{r}}\partial^{(3)}_{\tilde{c}\tilde{e}\tilde{0}}d_{ab}
\partial^{(3)}_{\tilde{d}\tilde{r}\tilde{0}}d_{ab}]\cdot
 e^{-2\lambda f_\epsilon}\eta_{\epsilon} dV_g
\\&+\sum_{c=0}^3[||\partial^{(2)}_{c0}d_{ab}
\cdot\eta_{\epsilon}||^2_{L^2_\lambda}+
||\partial_{c}d_{ab}\cdot\eta_{\epsilon}||^2_{L^2_\lambda}+
||\partial^{(3)}_{c00}d_{ab}\cdot\eta_{\epsilon}||^2_{L^2_\lambda}].
\end{split}
\end{equation}

Then, performing the same integration by parts as for (\ref{prfstim2}) we finally derive the
analogue of (\ref{prfstim3}):

 \begin{equation}
\label{prfstim3'}
\begin{split}
&\sum_{a,b=0}^3||(\partial_0g^{yu})\partial^{(3)}_{yu0}d_{ab}\cdot\eta_{\epsilon}
||_{L^2_\lambda}\le C'\{\sum_{a,b=0}^3||
\widetilde{\Box_g}\partial_{0}d_{ab}\cdot
 \eta_{\epsilon}||_{L^2_\lambda}+
\\& +\sum_{c=0}^3[||\partial^{(2)}_{c0}d_{ab}
\cdot\eta_{\epsilon}||_{L^2_\lambda}+
||\partial_{c}d_{ab}\cdot\eta_{\epsilon}||_{L^2_\lambda}+
||\partial^{(3)}_{c00}d_{ab}\cdot\eta_{\epsilon}||_{L^2_\lambda}]
+||V_{\epsilon}||_{L^2_\lambda}\}.
\end{split}
\end{equation}
Then, again invoking Lemma \ref{easylem} we derive our claim. $\Box$
\newline

{\bf Proof of (\ref{suid2}):} Again we start by applying Lemma \ref{derivlem},
and then use the equation (\ref{ss5}):

\begin{equation}
\label{karaiskakh}
\begin{split}
&\sum_{a,b=0}^3||D_{[ab]}\cdot\eta_{\epsilon}
||_{L^2_\lambda}\le \frac{C}{\sqrt{\lambda}}\sum_{a,b=0}^3||\partial_0D_{[ab]}\cdot\eta_{\epsilon}
||_{L^2_\lambda}+||V_{\epsilon}||_{L^2_\lambda}\le
\\& \frac{C}{\sqrt{\lambda}}\{\sum_{a,b=0}^3||d_{ab}\cdot\eta_{\epsilon}
||_{L^2_\lambda}+
\sum_{a,b,c=0}^3||\partial_cd_{ab}\cdot \eta_{\epsilon}||_{L^2_\lambda}+
\sum_{a,b,c,d=0}^3||T_{abcd}\cdot\eta_{\epsilon}||_{L^2_\lambda}+
\sum_{a,b,c=0}^3||\partial_{0c}d_{ab}\cdot\eta_{\epsilon}||_{L^2_\lambda}+
\\&\sum_{a,b=0}^3||D_{ab}\cdot\eta_{\epsilon}||_{L^2_\lambda}+
\sum_{a,b=0}^3||\sum_{i,j=1}^2c^{ij}\partial_iG_{j[a,b]}\cdot\eta_{\epsilon}
||_{L^2_\lambda}+||V_{\epsilon}||_{L^2_\lambda}\}.
\end{split}
\end{equation}

Now, observe that all the terms in the second line can be controlled by virtue of Lemma \ref{easylem}.
Furthermore, we straightforwardly obtain:

$$\sum_{a,b=0}^3||D_{ab}||_{L^2_\lambda}\le
\sum_{a,b=0}^3||D_{(ab)}||_{L^2_\lambda}+
\sum_{a,b=0}^3||D_{[ab]}||_{L^2_\lambda},$$
and now the term $\sum_{a,b=0}^3||D_{[ab]}||_{L^2_\lambda}$
can be absorbed into the LHS (when $\lambda$ is large enough), while 
we have already controlled $\sum_{a,b=0}^3||D_{(ab)}||_{L^2_\lambda}$
by virtue of equation (\ref{suid1}).
Thus, matters are reduced to controlling the term
$\sum_{a,b=0}^3||\sum_{i,j=1}^2c^{ij}\partial_iG_{j[a,b]}||_{L^2_\lambda}$.

\par To do this we again will use the vector fields $\tilde{X}^1,\tilde{X}^2$
defined above (see the discussion after (\ref{rough1})), and we will evaluate the tensor ${\bf G_{ab,c}}$ 
against those vector fields.

\par Using this observation we can again derive that:

$$|\sum_{i,j=1}^2c^{ij}\partial_iG_{j[a,b]}|^2\le C\cdot
|\sum_{\tilde{i},\tilde{j}=1}^2 \sum_{\tilde{z},\tilde{q}=0}^2 g^{\tilde{i}\tilde{j}}g^{\tilde{z}\tilde{q}}
\partial_{\tilde{i}}G_{\tilde{z}[a,b]}\partial_{\tilde{j}}G_{\tilde{q}[a,b]}+ \sum_{c=0}^3[|\partial_0G_{c[a,b]}|^2+|\partial_cG_{0[a,b]}|^2]. $$
Now, we straightforwardly see  that
$\sum_{a,b,c=0}^3|\partial_0G_{c[a,b]}|+|\partial_cG_{0[a,b]}|\le
C\sum_{a,b,c=0}^3|\partial^{(2)}_{0c}d_{ab}|$. Thus matters are reduced to controlling
$\sum_{a,b=0}^3\int_{B_{\epsilon^{10}}}\sum_{\tilde{i},\tilde{j}=1}^2
\sum_{\tilde{z},\tilde{q}=0}^2 g^{\tilde{i}\tilde{j}}g^{\tilde{z}\tilde{q}}
\partial_{\tilde{i}}G_{\tilde{z}[a,b]}\partial_{\tilde{j}}
G_{\tilde{q}[a,b]}\cdot \eta^2_{\epsilon}dV_g$. We will prove:

\begin{equation}
 \label{theend}
\begin{split}
& \sum_{a,b=0}^3\int_{B_{\epsilon^{10}}}\sum_{\tilde{i},\tilde{j}=1}^2
\sum_{\tilde{z},\tilde{q}=0}^2 g^{\tilde{i}\tilde{j}}g^{\tilde{z}\tilde{q}}
\partial_{\tilde{i}}G_{\tilde{z}[a,b]}\partial_{\tilde{j}}
G_{\tilde{q}[a,b]}\cdot \eta^2_{\epsilon}dV_g\le
\\& C\{ \sum_{a,b=0}^3||D_{[ab]}\cdot \eta_{\epsilon}||^2_{L^2_\lambda}+
\sum_{a,b,c=0}^3||\partial_cd_{[ab]}\cdot \eta_{\epsilon}||^2_{L^2_\lambda}+
\sum_{a,b,c=0}^3||\partial^{(2)}_{0c}d_{[ab]}\cdot \eta_{\epsilon}||^2_{L^2_\lambda}\}.
\end{split}
\end{equation}

We do this by first commuting indices using the relation:

 $$\partial_{\tilde{\alpha}}G_{\tilde{\beta} i,j}- \partial_{\tilde{\beta}}G_{\tilde{\alpha} i,j}=
 T_{\tilde{\alpha}\tilde{\beta}ij}+L_{\tilde{\alpha}\tilde{\beta}ij}^{yur}G_{yu,r}+
L'^{yu}_{\tilde{\alpha}\tilde{\beta}ij}d_{yu},$$
and then integrating by parts. Explicitly we derive:

\begin{equation}
\label{predeficit}
\begin{split}
&\sum_{a,b=0}^3\int_{B_{\epsilon^{10}}}\sum_{\tilde{i},\tilde{j}=1}^2 \sum_{\tilde{z},\tilde{q}=0}^2 g^{\tilde{i}\tilde{j}}g^{\tilde{z}\tilde{q}}
\partial_{\tilde{i}}G_{\tilde{z}[a,b]}\partial_{\tilde{j}}
G_{\tilde{q}[a,b]}e^{-2\lambda f_\epsilon}\cdot \eta^2_{\epsilon}dV_g=
\\&\sum_{a,b=0}^3\int_{B_{\epsilon^{10}}}
\sum_{\tilde{i},\tilde{j}=1}^2 \sum_{\tilde{z},\tilde{q}=0}^2
g^{\tilde{i}\tilde{j}}g^{\tilde{z}\tilde{q}}
\partial_{\tilde{z}}G_{\tilde{i}[a,b]}\partial_{\tilde{j}}
G_{\tilde{q}[a,b]}e^{-2\lambda f_\epsilon}\cdot \eta^2_{\epsilon}dV_g+
\\&\sum_{a,b=0}^3\int_{B_{\epsilon^{10}}}\sum_{\tilde{i},\tilde{j}=1}^2 \sum_{\tilde{z},\tilde{q}=0}^2 g^{\tilde{i}\tilde{j}}g^{\tilde{z}\tilde{q}}
T_{\tilde{i}\tilde{z} ab}\partial_{\tilde{j}}
G_{\tilde{q}[a,b]}e^{-2\lambda f_\epsilon}\cdot \eta^2_{\epsilon}dV_g+
\\&\sum_{a,b=0}^3\int_{B_{\epsilon^{10}}}\sum_{\tilde{i},\tilde{j}=1}^2 \sum_{\tilde{z},\tilde{q}=0}^2 g^{\tilde{i}\tilde{j}}g^{\tilde{z}\tilde{q}}
L_{\tilde{i}\tilde{z}ij}^{yur}G_{yu,r}
\partial_{\tilde{j}}G_{\tilde{q}[a,b]}e^{-2\lambda f_\epsilon}\cdot
 \eta^2_{\epsilon}dV_g+
 \\&\sum_{a,b=0}^3\int_{B_{\epsilon^{10}}}\sum_{\tilde{i},\tilde{j}=1}^2 \sum_{\tilde{z},\tilde{q}=0}^2 g^{\tilde{i}\tilde{j}}g^{\tilde{z}\tilde{q}}
L_{\tilde{i}\tilde{z}ij}^{yu}d_{yu}
\partial_{\tilde{j}}G_{\tilde{q}[a,b]}e^{-2\lambda f_\epsilon}\cdot
 \eta^2_{\epsilon}dV_g.
\end{split}
\end{equation}

To control the term in the second line of the RHS we apply 
the commutation relation again to derive:

\begin{equation}
\label{shmant}
\begin{split}
&\sum_{a,b=0}^3\int_{B_{\epsilon^{10}}}\sum_{\tilde{i},\tilde{j}=1}^2 \sum_{\tilde{z},\tilde{q}=0}^2 g^{\tilde{i}\tilde{j}}g^{\tilde{z}\tilde{q}}
T_{\tilde{i}\tilde{z} ab}\partial_{\tilde{j}}
G_{\tilde{q}[a,b]}e^{-2\lambda f_\epsilon}\cdot \eta^2_{\epsilon}dV_g=
\frac{1}{2}\sum_{a,b=0}^3\{\int_{B_{\epsilon^{10}}}\sum_{\tilde{i},\tilde{j}=1}^2
\sum_{\tilde{z},\tilde{q}=0}^2 g^{\tilde{i}\tilde{j}}g^{\tilde{z}\tilde{q}}
T_{\tilde{i}\tilde{z}ab}T_{\tilde{j}\tilde{q}ab}e^{-2\lambda f_\epsilon}\eta^2_\epsilon dV_g
\\&+ \int_{B_{\epsilon^{10}}}F^{yuqwzxv}_{ab}
T_{yuqw}G_{zx,v} e^{-2\lambda f_\epsilon}\cdot \eta^2_{\epsilon}dV_g +
\int_{B_{\epsilon^{10}}}F^{yuqwzx}_{ab}
T_{yuqw}d_{zx} e^{-2\lambda f_\epsilon}\cdot \eta^2_{\epsilon}dV_g\}
\le
\\&C\{ \sum_{a,b,c,d=0}^3||T_{abcd}
\cdot\eta_{\epsilon}||^2_{L^2_\lambda}+
\sum_{a,b,c=0}^3||G_{ab,c}
\cdot\eta_{\epsilon}||^2_{L^2_\lambda}+\sum_{a,b=0}^3||d_{ab}
\cdot\eta_{\epsilon}||^2_{L^2_\lambda}\}.
\end{split}
\end{equation}

Now to control the last two lines in the RHS of (\ref{predeficit}) 
we apply Cauchy-Schwartz to derive that for any $\rho>0$:

\begin{equation}
\label{shmant'}
\begin{split}
&\sum_{a,b=0}^3\int_{B_{\epsilon^{10}}}\sum_{\tilde{i},\tilde{j}=1}^2 \sum_{\tilde{z},\tilde{q}=0}^2 g^{\tilde{i}\tilde{j}}g^{\tilde{z}\tilde{q}}
L_{\tilde{i}\tilde{z}ij}^{yur}G_{yu,r}
\partial_{\tilde{j}}G_{\tilde{q}[a,b]}e^{-2\lambda f_\epsilon}\cdot
 \eta^2_{\epsilon}dV_g+
\\& \sum_{a,b=0}^3\int_{B_{\epsilon^{10}}}\sum_{\tilde{i},\tilde{j}=1}^2 \sum_{\tilde{z},\tilde{q}=0}^2 g^{\tilde{i}\tilde{j}}g^{\tilde{z}\tilde{q}}
L_{\tilde{i}\tilde{z}ij}^{yu}d_{yu}
\partial_{\tilde{j}}G_{\tilde{q}[a,b]}e^{-2\lambda f_\epsilon}\cdot
 \eta^2_{\epsilon}dV_g \le
\\&\frac{1}{\rho}\sum_{y,u,r=0}^3\int_{B_{\epsilon^{10}}}
(G_{yu,r})^2e^{-2\lambda f_\epsilon}\cdot \eta^2_{\epsilon}dV_g+
\frac{1}{\rho}\sum_{y,u=0}^3\int_{B_{\epsilon^{10}}}
(d_{yu})^2e^{-2\lambda f_\epsilon}\cdot \eta^2_{\epsilon}dV_g+
\\&C\cdot \rho\cdot \sum_{a,b=0}^3\int_{B_{\epsilon^{10}}}\sum_{\tilde{i},\tilde{j}=1}^2 \sum_{\tilde{z},\tilde{q}=0}^2 g^{\tilde{i}\tilde{j}}g^{\tilde{z}\tilde{q}}
\partial_{\tilde{i}}G_{\tilde{z}[a,b]}\partial_{\tilde{j}}
G_{\tilde{q}[a,b]}e^{-2\lambda f_\epsilon}\cdot \eta^2_{\epsilon}dV_g,
\end{split}
\end{equation}
where the constant $C$ is universal (meaning that it does not depend on
$\lambda$ or $\rho$--it merely depends on the norms of the tensors $g^{ab}$ etc).
We then replace the two equations above into (\ref{predeficit}).
Picking $\rho$ small enough we can absorb the term $\sum_{\tilde{z},\tilde{q}=0}^2 g^{\tilde{i}\tilde{j}}g^{\tilde{z}\tilde{q}}
\partial_{\tilde{i}}G_{\tilde{z}[a,b]}\partial_{\tilde{q}}
G_{\tilde{q}[a,b]}e^{-2\lambda f_\epsilon}\cdot \eta^2_{\epsilon}dV_g$
in the above into the LHS
of (\ref{predeficit}). Thus, matters are reduced to controlling the term $\sum_{a,b=0}^3\int_{B_{\epsilon^{10}}}\sum_{\tilde{i},\tilde{j}=1}^2
\sum_{\tilde{z},\tilde{q}=0}^2 g^{\tilde{i}\tilde{j}}g^{\tilde{z}\tilde{q}}
\partial_{\tilde{z}}G_{\tilde{i}[a,b]}\partial_{\tilde{j}}
G_{\tilde{q}[a,b]}e^{-2\lambda f_\epsilon}\cdot \eta^2_{\epsilon}dV_g$
in the RHS of (\ref{predeficit}). This can be done by
integrating by parts twice, first the derivative $\partial_{\tilde{z}}$
and then the derivative $\partial_{\tilde{j}}$:\footnote{Since the vector fields 
$\partial_{\tilde{e}}, {}_{\tilde{e}}={}_{\tilde{1}},{}_{\tilde{2}}$  are tangent 
to the level sets of $f_\epsilon$,
we do not bring out derivatives of the factor $e^{-\lambda f_\epsilon}$.}

\begin{equation}
\label{byparts}
\begin{split}
&\sum_{a,b=0}^3\int_{B_{\epsilon^{10}}}\sum_{\tilde{i},\tilde{j}=1}^2 \sum_{\tilde{z},\tilde{q}=0}^2 g^{\tilde{i}\tilde{j}}g^{\tilde{z}\tilde{q}}
\partial_{\tilde{z}}G_{\tilde{i}[a,b]}\partial_{\tilde{j}}
G_{\tilde{q}[a,b]}e^{-2\lambda f_\epsilon}\cdot \eta^2_{\epsilon}dV_g=
\int_{B_{\epsilon^{10}}}V_{\epsilon}e^{-2\lambda f_\epsilon}dV_g+
\\&\sum_{a,b=0}^3\int_{B_{\epsilon^{10}}}\sum_{\tilde{i},\tilde{j}=1}^2 \sum_{\tilde{z},\tilde{q}=0}^2 g^{\tilde{i}\tilde{j}}g^{\tilde{z}\tilde{q}}
\partial_{\tilde{j}}G_{\tilde{i}[a,b]}\partial_{\tilde{z}}
G_{\tilde{q}[a,b]}e^{-2\lambda f_\epsilon}\cdot \eta^2_{\epsilon}dV_g+
\\&\sum_{a,b=0}^3\int_{B_{\epsilon^{10}}}g^{\tilde{i}\tilde{j}}L_{ab}^{yur}G_{yu,r}
\partial_{\tilde{i}}G_{\tilde{j}a,b}e^{-2\lambda f_\epsilon}\cdot \eta^2_{\epsilon}dV_g+
\sum_{a,b=0}^3\int_{B_{\epsilon^{10}}}L_{ab}^{yur}G_{yu,r}L'^{yur}_{ab}G_{yu,r}
e^{-2\lambda f_\epsilon}\cdot \eta^2_{\epsilon}dV_g.
\end{split}
\end{equation}

Applying Cauchy-Schwartz to the expressions
$\sum_{a,b=0}^3\int_{B_{\epsilon^{10}}}g^{\tilde{i}\tilde{j}}L_{ab}^{yur}G_{yu,r}
\partial_{\tilde{i}}G_{\tilde{j}a,b}e^{-2\lambda f_\epsilon}\cdot \eta^2_{\epsilon}dV_g$ and  $\sum_{a,b=0}^3\int_{B_{\epsilon^{10}}}g^{\tilde{i}\tilde{j}}L_{ab}^{yur}G_{yu,r}
\partial_{\tilde{i}}G_{\tilde{j}a,b}e^{-2\lambda f_\epsilon}\cdot \eta^2_{\epsilon}dV_g$
and then replacing (\ref{byparts}), (\ref{shmant}), (\ref{shmant'})
into (\ref{predeficit}) we derive:

\begin{equation}
\label{deficit}
\begin{split}
&\sum_{a,b=0}^3\int_{B_{\epsilon^{10}}}
\sum_{\tilde{i},\tilde{j},\tilde{z},\tilde{q}=\tilde{1}}^{\tilde{2}}
 g^{\tilde{i}\tilde{j}}g^{\tilde{z}\tilde{q}}
\partial_{\tilde{i}}G_{\tilde{z}[a,b]}\partial_{\tilde{j}}
G_{\tilde{q}[a,b]}e^{-2\lambda f_\epsilon}\cdot \eta^2_{\epsilon}dV_g\le
\\& \sum_{a,b=0}^3\int_{B_{\epsilon^{10}}}|\sum_{\tilde{i},\tilde{j},\tilde{z},\tilde{q}=1}^2  g^{\tilde{i}\tilde{j}}g^{\tilde{z}\tilde{q}}
\partial_{\tilde{i}}G_{\tilde{j}[a,b]}\partial_{\tilde{z}}
G_{\tilde{q}[a,b]}e^{-2\lambda f_\epsilon}\cdot \eta^2_{\epsilon}dV_g+
\sum_{a,b,c,d=0}^3||T_{abcd}\cdot\eta_{\epsilon}||_{L^2_\lambda}+
\\& \sum_{a,b,c=0}^3||\partial_{c}d_{ab} ||_{L^2_\lambda} 
+\sum_{a,b,c=0}^3||\partial^{(2)}_{c0}d_{ab} ||_{L^2_\lambda}.
\end{split}
\end{equation}
Finally we again use the fact that $g^{\tilde{3}\tilde{a}}=0$ for ${}^{\tilde{a}}={}^1,{}^2,{}^3$ 
to derive:

\begin{equation}
\label{deficit'}
\begin{split}
&\sum_{a,b=0}^3\int_{B_{\epsilon^{10}}}
\sum_{\tilde{i},\tilde{j},\tilde{z},\tilde{q}=\tilde{1}}^{\tilde{2}}  g^{\tilde{i}\tilde{j}}g^{\tilde{z}\tilde{q}}
\partial_{\tilde{i}}G_{\tilde{j}[a,b]}\partial_{\tilde{z}}
G_{\tilde{q}[a,b]}\cdot e^{-2\lambda f_\epsilon}\cdot
 \eta^2_{\epsilon}dV_g\le\sum_{a,b,c=0}^3||\partial_{c}d_{ab} ||^2_{L^2_\lambda} +
\\&\sum_{a,b=0}^3\int_{B_{\epsilon^{10}}}
\sum_{\tilde{i},\tilde{j},\tilde{z},\tilde{q}=\tilde{0}}^{\tilde{3}}
 g^{\tilde{i}\tilde{j}}g^{\tilde{z}\tilde{q}}
\partial_{\tilde{i}}G_{\tilde{j}[a,b]}\partial_{\tilde{z}}
G_{\tilde{q}[a,b]}\cdot e^{-2\lambda f_\epsilon}\cdot\eta^2_{\epsilon}dV_g
+\sum_{a,b,c=0}^3||\partial^{(2)}_{c0}d_{ab} ||^2_{L^2_\lambda}.
\end{split}
\end{equation}
Since $g^{\tilde{i}\tilde{j}}
\partial_{\tilde{i}}G_{\tilde{j}[a,b]}\partial_{\tilde{z}}=D_{[ab]}$, combining this
with (\ref{deficit}) we derive (\ref{theend}) and thus our claim. $\Box$

\section{A derivation of Theorem \ref{thetheorem} under minimal hypotheses.}
\label{power}

We now show that the conclusion of Theorem \ref{thetheorem} can in fact
be derived under much weaker assumptions. We will see that for two vacuum
space-times $(M,{\bf g}), (\tilde{M},\tilde{\bf g})$ with horizons
${\cal H}^{+}\bigcup {\cal H}^{-}$, ${\cal \tilde{H}}^{+}\bigcup {\cal \tilde{H}}^{-}$
 to be isometric in some open neighborhoods of $P\in S,\tilde{P}\in \tilde{S}$, it suffices
 to assume that (near $P,\tilde{P}$) the conformal structures induced by ${\bf g}, \tilde{\bf g}$ onto
 $({\cal H}^{+}\bigcup {\cal H}^{-}), ({\cal \tilde{H}}^{+}\bigcup {\cal \tilde{H}}^{-})$ are 
 equivalent,\footnote{This notion will be made precise below.} that
 the spheres $S,\tilde{S}$ with their induced metrics from ${\bf g}, \tilde{\bf g}$ are isometric and moreover that
the second fundamental forms of the spheres $S,\tilde{S}$ also agree.
The proof of this claim will rely only on an analysis of the Taylor series expansion of the metrics
${\bf g}, \tilde{\bf g}$ on the bifurcate horizons, followed by an application of
Theorem \ref{thetheorem}. It is worth noting that this analysis of {\it free data} 
is in complete agreement with the work of Rendall in \cite{rendall}.
Nonetheless we are not able to invoke his work directly, since \cite{rendall} considers the
metric expressed in wave coordinates, as opposed to our double Fermi coordinates.
\newline

In order to introduce the weaker requirements needed for our stronger version of Theorem 
\ref{thetheorem} we define:

\begin{definition}
\label{strongconf}
We say that the two space-times $(M,{\bf g}), (\tilde{M}, {\bf \tilde{g}})$ with
horizons ${\cal H}^{+}\bigcup {\cal H}^{-}$, ${\cal \tilde{H}}^{+}\bigcup {\cal \tilde{H}}^{-}$
are weakly equivalent near points $P\in S={\cal H}^{+}\bigcap {\cal H}^{-}$,
$\tilde{P}\in \tilde{S}={\cal \tilde{H}}^{+}\bigcap {\cal \tilde{H}}^{-}$ if there exist two double Fermi coordinate
systems ${\cal Y}, \tilde{\cal Y}$ for $M_1, M_2$ in open 
neighborhoods $\Omega,\tilde{\Omega}$ of the points $P,\tilde{P}$ so that
the map $\Phi=\tilde{\cal Y}\circ {\cal Y}^{-1}$ satisfies the following properties:

\begin{enumerate}

\item{There exists a function $f\in C^{4}({\cal H}^{+}\bigcup {\cal H}^{-})$ so that
$\Phi^{*}(\tilde{\bf g}|_{{\cal \tilde{H}}^{+}\bigcup {\cal \tilde{H}}^{-}\bigcap \tilde{\Omega}})=e^{2f} 
{\bf g}|_{{\cal H}^{+}\bigcup {\cal H}^{-}\bigcap \Omega}$.}

\item{On $S\bigcap \Omega$ we have $f=\frac{\partial}{\partial x^0}f=\frac{\partial}{\partial x^3}f=0$.}

\item{On $S\bigcap \Omega$ we also have $k_{13}=k'_{13}$, $k_{23}=k'_{23}$, where ${\bf k}$, ${\bf k'}$
stand for the second fundamental forms of ${\cal H}^{+}$ for the metrics
${\bf g}, \Phi^{*}{\bf g}$ with respect to the null vector field $\frac{\partial}{\partial x^3}$.}
\end{enumerate}
\end{definition}

Our strengthened theorem is then the following:

\begin{theorem}
\label{strength}
Consider two vacuum space-times $(M,{\bf g}), (\tilde{M},\tilde{\bf g})$
as in the discussion above Theorem \ref{thetheorem}, with horizons
$({\cal H}^{+}\bigcup {\cal H}^{-})$, $({\cal \tilde{H}}^{+}\bigcup {\cal \tilde{H}}^{-})$.

Then if these two horizons are weakly equivalent near the points $P,\tilde{P}$, then 
the map $\Phi$ in the definition above is an isometry, when restricted to a 
small enough open neighborhood of  $P,\tilde{P}$.
\end{theorem}

{\it Proof of Theorem \ref{strength}:} We only need to show that $\Phi^{*}\tilde{\bf g}- {\bf g}$
vanishes to third order on $({\cal H}^{+}\bigcup {\cal H}^{-})\bigcap \Omega$; Theorem \ref{thetheorem} will then imply
that $(M,{\bf g}), (\tilde{M},\tilde{\bf g})$
are isometric in open neighborhoods of $P,\tilde{P}$. We first prove this claim on ${\cal H}^{+}$.

{\it Proof of the claim on ${\cal H}^{+}$:} 
We will show this in two steps:
Firstly we will prove that $f=0$ on ${\cal H}^{+}$. Then we will show that the jets up
to third order of ${\bf g}$, $\Phi^{*}\tilde{\bf g}$ are uniquely determined
 given the values of $k_{13},k_{23}$,
$k'_{13},k'_{23}$. Given the hypothesis of our Lemma, that will
prove that $\Phi^{*}\tilde{\bf g}-{\bf g}$ vanishes to third order on ${\cal H}^{+}$.
 We introduce a notational
 convention to simplify our task: At each stage of our proof we will denote by
 $K(x^1,x^2,x^3), K'(x^1,x^2,x^3), \dots$ (or just $K,K',\dots$ for short) 
 generic {\it known} functions defined over ${\cal H}^{+}$.

 We first show that $f=0$ on ${\cal H}^{+}$: Consider the equation
 $Ric_{33}(\Phi^{*}{\bf \tilde{g}})-Ric_{33}({\bf g})=0$ on ${\cal H}^{+}$.
 We thus derive an equation:

 $$\partial^{(2)}_{33} f-(\partial_3f)^2=K(x^1,x^2,x^3)\partial_3f$$
Thus, given that $f=\partial_0f=0$ in $S$, we derive that $f=0$ by the fundamental theorem of ODEs.

 Now, we recall that
 $g_{00}=\tilde{g}_{00}=g_{01}=\tilde{g}_{01}=g_{02}=\tilde{g}_{02}=0$
 throughout $\Omega$. Furthermore, since the integral curves of
 $\frac{\partial }{\partial x^3}$ in ${\cal H}^{+}_1$ are null geodesics,
 and the fact that $g_{33}=g_{13}=g_{23}=0$ on $S$, we derive that
 $g_{33}=g_{13}=g_{23}=0$ throughout ${\cal H}^{+}$. Finally, since integral curves are 
arc-length paramatrized geodesics, we derive that
 $\partial_0g_{33}=0$ throughout ${\cal H}^{+}$. (The same relations are of
 course true for the corresponding components of $(\Phi^{*}\tilde{g})$).

Now we will show that the components
 $\partial^{(k)}_{0\dots 0}g_{13},\partial^{(k)}_{0\dots 0}g_{23}$,
 $\partial^{(k)}_{0\dots 0}g_{11},\partial^{(k)}_{0\dots 0}g_{12}$,
 $\partial^{(k)}_{0\dots 0}g_{22}$ for $1\le k\le 2$,
 and $\partial^{(2)}_{00}g_{33}$ on ${\cal H}^{+}$ can be uniquely determined
 by the above relations given the equation $Ric(g)=0$ and also the data $\partial_0{ g}_{13}$,
$\partial_0 {g}_{23}$ on $S$. This will {\it also} show that the corresponding components of
$\Phi^{*}\tilde{g}, \partial(\Phi^{*}\tilde{g}), \partial^{(2)}(\Phi^{*}\tilde{g})$
are uniquely determined by the equation $Ric(\Phi^{*}\tilde{g})=0$ and the $\partial_0(\Phi^{*}\tilde{g})_{13}$,
$\partial_0 (\Phi^{*}\tilde{g})_{23}$ on $S$ and we will thus derive our claim on ${\cal H}^{+}$.

We determine the above components of the jet of $g$ in stages: Firstly observe that in the notation of Definition
\ref{strongconf}, for any point on $S$: $k_{13}=\frac{1}{2}\partial_0g_{13}$, $k'_{13}=\frac{1}{2}\partial_0(\Phi^{*}\tilde{g}_{13})$
and also $k_{23}=\frac{1}{2}\partial_0g_{23}$, $k'_{23}=\frac{1}{2}\partial_0(\Phi^{*}\tilde{g}_{23})$.
Next, we will derive ODEs involving the unknowns $\partial_0 g_{13}, \partial_0 g_{23}$ in ${\cal H}^{+}$:
 Consider the equations $Ric_{13}=0, Ric_{23}=0$. We derive:

 $$\partial_3(\partial_0g_{13}) =K(x^1,x^2,x^3),
\partial_3(\partial_0g_{23}) =K(x^1,x^2,x^3).$$
Thus, again by the fundamental theorem of ODEs we determine the functions
$\partial_0g_{13},\partial_0g_{23}$ on ${\cal H}^{+}$.
Now we use the equations $Ric_{11}=0, Ric_{12}=0, Ric_{22}=0$ to derive:

\begin{eqnarray}
\label{mikhs1}
&\partial_3(\partial_0g_{11}) =K\partial_0 g_{11}+K'\partial_0 g_{12}+
K''\partial_0 g_{22}+K''',
\label{mikhs2}
\\&\partial_3(\partial_0g_{12}) =K(x^1,x^2,x^3)\partial_0 g_{11}+K'\partial_0 g_{12}+
K''\partial_0 g_{22}+K''',
\label{mikhs3}
\\&\partial_3(\partial_0g_{22}) =K\partial_0 g_{11}+K'\partial_0 g_{12}+
K''\partial_0 g_{22}+K'''.
\end{eqnarray}

Thus we again invoke the fundamental theorem of ODEs to conclude that the functions $\partial_0g_{11},\partial_0g_{12},\partial_0g_{22}$ are uniquely determined on ${\cal H}^{+}$
by the  data we have prescribed.

 Similarly, considering the equation $Ric_{03}=0$ on ${\cal H}^{+}$ we calculate 
$\partial^{(2)}_{00}g_{33}$ and then using the equations $Ric_{01}=Ric_{02}=0$ we calculate $\partial^{(2)}_{00}g_{13}, \partial^{(2)}_{00}g_{13}$ on ${\cal H}^{+}$.
Finally, considering the equations
$\partial_0Ric_{11}=0, \partial_0Ric_{12}=0,\partial_0Ric_{22}=0$ and
repeating the argument we used for the equations (\ref{mikhs1}), (\ref{mikhs2}), (\ref{mikhs3})
we calculate $\partial^{(2)}_{00}g_{11}, \partial^{(2)}_{00}g_{12},\partial^{(2)}_{00}g_{22}$. 
This proves our claim on ${\cal H}^{+}$.
\newline

{\it Proof of the claim on ${\cal H}^{-}$:} Again, we only need to show that $f=0$ 
on ${\cal H}^{-}$ and that
 the 2-jets of $g$ are uniquely determined by the equation $Ric(g)=0$
 and the data $\partial_0(g_{13}$), $\partial_0(g_{23})$ on $S$.

Again, at each stage of our proof we will denote by
$K(x^0,x^1,x^2), K'(x^0,x^1,x^2), \dots$ (or just $K,K',\dots$ for short)
generic {\it known} functions defined over ${\cal H}^{-}$.

To show that $f=0$ on ${\cal H}^{-}$ we
consider the equation $Ric_{00}(\Phi^{*}\tilde{g})-Ric_{00}(g)=0$ on ${\cal H}^{-}$.
We derive an ODE on the conformal factor $f$:

$$\partial^{(2)}_{00}f+(Const)\cdot (\partial_0 f)^2+K\partial_0f=0 $$
Thus we derive that $f=0$ on ${\cal H}^{-}$ using the fundamental theorem of ODE and the hypothesis that
 $f=\partial_3f=0$ on $S$.

Now, the unknowns we wish to determine are the components $\partial^{(k)}_{3\dots 3}g_{13}$,
$\partial^{(k)}_{3\dots 3}g_{23},\partial^{(k)}_{3\dots 3}g_{33}$ for every $k, 0\le k\le 2$ and the components
$\partial^{(k)}_{3\dots 3}g_{11}$, $\partial^{(k)}_{3\dots 3}g_{12}$, $\partial^{(k)}_{3\dots 3}g_{22}$
for every $k, 1\le k\le 2$.

Using the equations $Ric_{01}=0$, $Ric_{02}=0$ we derive:

$$\partial^{(2)}_{00}g_{13}+K\partial_0g_{13}+K'g_{23}=K'',
\partial^{(2)}_{00}g_{23}+\tilde{K}\partial_0g_{13}+\tilde{K}'g_{23}=\tilde{K}''.$$
Therefore by applying the fundamental theorem of ODEs to the above 
two equations we derive that we can determine $g_{13}, g_{23}$ from the initial data on $S$.
Now we also consider the equations $Ric_{11}=0, Ric_{12}=0, Ric_{22}=0$. We derive a system of three equations:

\begin{eqnarray}
&\partial_0(\partial_3g_{11})+K_1 \partial_3g_{11} +K_2 \partial_3g_{12}+K_3 \partial_3g_{22}=K_4,
\\&\partial_0(\partial_3g_{12})+\tilde{K}_1 \partial_3g_{11} +\tilde{K}_2 \partial_3g_{12}+\tilde{K}_3 \partial_3g_{22}=\tilde{K}_4,
\\&\partial_0(\partial_3g_{22})+\overline{K}_1 \partial_3g_{11} +\overline{K}_2 \partial_3g_{12}+\overline{K}_3 \partial_3g_{22}=\overline{K}_4.
\end{eqnarray}

Thus, we also derive that the values of $\partial_0g_{11},\partial_0g_{12},\partial_0g_{22}$
can be determined on ${\cal H}^{-}$ from our prescribed data.

Finally, using the equation $Ric_{03}=0$ on ${\cal H}^{-}$ we derive an equation:

$$\partial^{(2)}_{00}g_{33}+K_1\partial_0g_{33}=K_2. $$
Therefore, using the fact that $\partial_0g_{33}=0$ on $S$ we derive that $g_{33}$
can also be determined on ${\cal H}^{-}$. Finally considering $\partial_3$-derivatives
of the equations  above, we can also 
determine all the higher derivatives of the unknown functions. $\Box$

\end{document}